\documentclass[aps,pra,floatfix,showpacs,preprint,superscriptaddress]{revtex4-1}
\usepackage{bm}
\usepackage{mathrsfs}
\usepackage{amsmath}
\usepackage{amssymb}
\usepackage{revsymb}
\usepackage{accents}
\usepackage{color}
\usepackage[force]{feynmp-auto}
\usepackage{graphicx}

\begin{document}
\title{One-loop vertex correction in a plane wave}
\author{A. \surname{Di Piazza}}
\email{dipiazza@mpi-hd.mpg.de}
\affiliation{Max Planck Institute for Nuclear Physics, Saupfercheckweg 1, D-69117 Heidelberg, Germany}
\author{M. A. \surname{Lopez-Lopez}}
\email{misha.lopez@umich.mx}
\affiliation{Instituto de F\'{i}sica y Matem\'{a}ticas, Universidad Michoacana de San Nicol\'{a}s de Hidalgo, Edificio C-3, Apdo. Postal 2-82, C.P. 58040, Morelia, Michoac\'{a}n, Mexico}
\affiliation{Max Planck Institute for Nuclear Physics, Saupfercheckweg 1, D-69117 Heidelberg, Germany}
\date{\today}

\begin{abstract}
We compute the general expression of the one-loop vertex correction in an arbitrary plane-wave background field for the case of two on-shell external electrons and an off-shell external photon. The properties of the vertex corrections under gauge transformations of the plane-wave background field and of the radiation field are studied. Concerning the divergences of the vertex correction, the infrared one is cured by assigning a finite mass to the photon, whereas the ultraviolet one is shown to be renormalized exactly as in vacuum. Finally, the corresponding expression of the vertex correction within the locally-constant crossed field is also derived and the high-field asymptotic is shown to scale according to the Ritus-Narozhny conjecture.

  
\end{abstract}
 
\maketitle

\section{Introduction}

The predictions of QED agree with experiments with impressive accuracy (see, e.g., Refs. \cite{Hanneke_2008,Sturm_2011}). The great success of QED has called for testing this theory under more extreme conditions as, for example, those provided by intense background electromagnetic fields. An electromagnetic field is denoted as ``intense'' in the realm of QED if it is of the order of the so-called ``critical'' field of QED: $F_{cr}=m^2/|e|=1.3\times 10^{16}\;\text{V/cm}=4.4\times 10^{13}\;\text{G}$ (from now on we employ units with $\epsilon_0=\hbar=c=1$ and $m$ and $e<0$ denote the electron mass and charge, respectively) \cite{Landau_b_4_1982,Fradkin_b_1991,Dittrich_b_1985}. Importantly, the presence of intense background electromagnetic fields allows for testing QED on a sector where nonlinear effects with respect to the background field strongly affect physical processes and the dynamics of charged particles. This sector is somewhat alternative to the high-energy one successfully investigated via conventional accelerators and it thus can serve as an independent ground test of QED.

High-power optical lasers are becoming a suitable tool to test QED at critical field strengths, which correspond to laser intensities of the order of $10^{29}\;\text{W/cm$^2$}$. In fact, although available lasers have reached peak intensities $I_0$ of the order of $5.5\times 10^{22}\;\text{W/cm$^2$}$ \cite{Yoon_2019} and upcoming facilities aim at $I_0\sim 10^{23}\text{-}10^{24}\;\text{W/cm$^2$}$ \cite{APOLLON_10P,ELI,CoReLS,XCELS}, the Lorentz invariance of the theory implies that the effective laser field strength at which a process occurs is the one experienced by the charges in their rest frame \cite{Mitter_1975,Ritus_1985,Ehlotzky_2009,Reiss_2009,Di_Piazza_2012,
Dunne_2014}. Since the amplitude of the laser field is boosted by a factor of the order of the relativistic Lorentz factor of the charge, an electron for definiteness, one can see that the strong-field QED regime, in which the background strength is effectively of the order of $F_{cr}$, can be entered already at intensities of the order of $10^{23}\;\text{W/cm$^2$}$, if the laser field counterpropagates with respect to an electron/positron with energy of the order of $500\;\text{MeV}$. 

In order to test QED in the strong-field regime by means of intense optical fields, it is essential that both experiments and theoretical predictions are correspondingly accurate. However, as it is understandable, first experiments in this regime have so far been designed especially to show the occurrence of phenomena like nonlinear Compton scattering \cite{Bula_1996}, nonlinear Breit-Wheeler pair production \cite{Burke_1997,Bamber_1999}, and radiation reaction \cite{Cole_2018,Poder_2018}, without aiming at obtaining high-accuracy results. Correspondingly, on the theory side, the basic strong-field QED processes like nonlinear Compton scattering \cite{Goldman_1964,Nikishov_1964,Ritus_1985,Baier_b_1998,Ivanov_2004,Boca_2009,
Harvey_2009,Mackenroth_2010,Boca_2011,Mackenroth_2011,Seipt_2011,Seipt_2011b,
Dinu_2012,Krajewska_2012,Dinu_2013,Seipt_2013,
Krajewska_2014,Wistisen_2014,Harvey_2015,Seipt_2016,Seipt_2016b,Angioi_2016,
Harvey_2016b,Angioi_2018,Di_Piazza_2018_c,Dinu_2018,Alexandrov_2019,Di_Piazza_2019,Ilderton_2019_b} and nonlinear Breit-Wheeler pair production \cite{Reiss_1962,Nikishov_1964,Narozhny_2000,Roshchupkin_2001,Reiss_2009,
Heinzl_2010b,Mueller_2011b,Titov_2012,Nousch_2012,Krajewska_2013b,Jansen_2013,
Augustin_2014,Meuren_2015,Meuren_2016,Di_Piazza_2019,King_2020} have been studied in detail at tree level by approximating the laser field as a plane wave (see also the reviews \cite{Mitter_1975,Ritus_1985,Ehlotzky_2009,Reiss_2009,Di_Piazza_2012}). However, even under the plane-wave approximation, the radiative corrections of these processes have never been computed. The reason is that calculations including the effects of the external laser field exactly are significantly more complex than the corresponding calculations in vacuum. The standard technique, in fact, is to work within the so-called Furry picture \cite{Furry_1951}, where the electron-positron field is quantized in the presence of the background field \cite{Fradkin_b_1991,Landau_b_4_1982}. This requires that the Dirac equation can be solved analytically in the presence of the background field, which has been achieved in Ref. \cite{Volkov_1935} in the case of a plane wave (see also Ref. \cite{Landau_b_4_1982}), the corresponding states being known as Volkov states. An alternative, equivalent technique is the so-called operator technique, first proposed by Schwinger \cite{Schwinger_1951} and then developed for the case of a background plane wave \cite{Baier_1976_a,Baier_1976_b,Di_Piazza_2007,Di_Piazza_2008_b,Di_Piazza_2013,
Di_Piazza_2018_d}, which does not require the explicit solution of the Dirac equation in the plane-wave field. 

\begin{figure}
\begin{center}
\includegraphics[width=0.6\columnwidth]{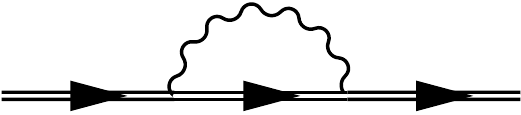}
\caption{The one-loop mass operator in an intense plane wave. The double lines represent exact electron states and propagator in a plane wave (Volkov states and propagator, respectively) \cite{Landau_b_4_1982}.}
\label{FD_MO}
\end{center}
\end{figure}

\begin{figure}
\begin{center}
\includegraphics[width=0.6\columnwidth]{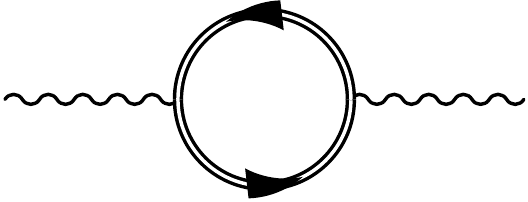}
\caption{The one-loop polarization operator in an intense plane wave. The double lines represent exact electron propagators in a plane wave (Volkov propagators) \cite{Landau_b_4_1982}.}
\label{FD_PO}
\end{center}
\end{figure}

\begin{figure}
\begin{center}
\includegraphics[width=0.4\columnwidth]{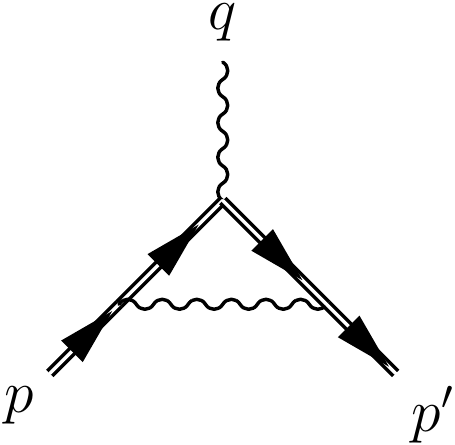}
\caption{The one-loop Feynman diagram corresponding to the vertex correction. The double lines represent exact electron states and propagator in a plane wave (Volkov states and propagator, respectively) \cite{Landau_b_4_1982}.}
\label{FD_VC}
\end{center}
\end{figure}

Going back to the radiative corrections, a systematic study has been only carried out in the special case of a zero-frequency plane wave or a constant crossed field, i.e., a constant and uniform electromagnetic field with electric and magnetic field having the same amplitude and being perpendicular to each other, from the early works of Ritus and Narozhny \cite{Ritus_1970,Ritus_1972,Narozhny_1979,Narozhny_1980,Morozov_1981} to the more recent one \cite{Mironov_2020} (see also Ref. \cite{Akhmedov_1983} and the reviews in Refs. \cite{Akhmedov_2011,Fedotov_2017}), where higher-loop Feynman diagrams have been evaluated. However, so far, in the case of a general plane wave with an arbitrary polarization and shape, only the one-loop mass operator (see Fig. \ref{FD_MO}) and the one-loop polarization operator (see Fig. \ref{FD_PO}) have been computed in Ref. \cite{Baier_1976_a} and in Refs. \cite{Becker_1975,Baier_1976_b}, respectively (see also Ref. \cite{Meuren_2013} for an alternative derivation of the polarization operator).
The one-loop vertex correction in a general plane wave (see Fig. \ref{FD_VC}) has never been evaluated, whereas the corresponding quantity in a constant crossed field was computed in Ref. \cite{Morozov_1981}.
The purpose of the present paper is to fill this gap and, indeed, to compute the one-loop vertex correction in an arbitrary plane wave for the case of two on-shell external electrons and an off-shell external photon. It is worth mentioning here that the computation of the vertex-correction function is not only important to evaluate the leading-order radiative corrections of strong-field QED processes. There is also a more fundamental reason related to the so-called Ritus-Narozhny conjecture \cite{Ritus_1970,Narozhny_1979,Narozhny_1980,Morozov_1981} about the high-energy behavior of radiative corrections in strong-field QED in a constant crossed field. As we have mentioned, a constant crossed field is a constant and uniform electromagnetic field $F_0^{\mu\nu}=(\bm{E}_0,\bm{B}_0)$ such that the two field Lorentz-invariants $\bm{E}_0^2-\bm{B}_0^2$ and $\bm{E}_0\cdot\bm{B}_0$ vanish. Now, in a constant crossed field radiative corrections depend only on the Lorentz- and gauge-invariant quantum nonlinearity parameter $\chi_0=\sqrt{-(p_{\mu}F_0^{\mu\nu})^2}/mF_{cr}$ \cite{Mitter_1975,Ritus_1985,Ehlotzky_2009,Reiss_2009,Di_Piazza_2012}, where $p^{\mu}$ is the four-momentum of the particle at hand and where the metric tensor $\eta^{\mu\nu}=\text{diag}(+1,-1,-1,-1)$ is employed. The Ritus-Narozhny conjecture states that at $\chi_0\gg 1$ the effective coupling of QED in a constant crossed field scales as $\alpha\chi_0^{2/3}$. Since, apart from irrelevant prefactors, the energy of the particle enters radiative corrections only through $\chi_0$ at $\chi_0\gg 1$, the Ritus-Narozhny conjecture implies an asymptotic high-energy behavior of strong-field QED in a constant crossed field qualitatively different from the logarithmic one of QED in vacuum \cite{Jauch_b_1976,Itzykson_b_1980,Landau_b_4_1982,Schwartz_b_2014}. The physical relevance of the Ritus-Narozhny conjecture is broadened by the so-called locally-constant field approximation (LCFA), stating that in the limit of low-frequency plane waves the probabilities of QED processes reduce to the corresponding probabilities in a constant crossed field averaged over the phase-dependent plane-wave profile \cite{Ritus_1985}. In Ref. \cite{Podszus_2019} we have investigated the one-loop mass and polarization operator to show that, if one first performs in the general expression of these quantities the high-energy limit, one indeed recovers the typical logarithmic behavior of QED as in vacuum (see also Ref. \cite{Ilderton_2019}). Below, we will also investigate the vertex correction within the LCFA, whereas the high-energy asymptotic will be presented elsewhere.

The paper is organized as follows: In Sec. \ref{Notation} we introduce the basic notation of the paper. In Sec. \ref{VC_General} the general form of the vertex-correction function is derived by means of the operator technique. In Sec. \ref{VC_GI} the properties of the vertex-correction function under gauge transformations of the radiation field and of the plane-wave background field are studied. In Sec. \ref{VC_CP} we show how to regularize and renormalize the vertex-correction function in the ultraviolet. The expression of the vertex-correction function within the LCFA is derived in Sec. \ref{VC_LCFA} and, finally, the main conclusions of the paper are reported in Sec. \ref{VC_Conclusions}. An appendix contains some technical considerations on a component of the vertex-correction function.

\section{Notation}
\label{Notation}
The notation employed below is the same as in Ref.~\cite{Di_Piazza_2018_d} but it is convenient to report here the main definitions. As we have mentioned in the Introduction, the present paper focuses on studying radiative corrections in a general plane-wave field. The latter is described by the four-vector potential $A^{\mu}(\phi)$, which only depends on the light-cone time $\phi=t-\bm{n}\cdot \bm{x}$. Here, the unit vector $\bm{n}$ defines the propagation direction of the plane wave, which can be used to introduce two useful four-dimensional quantities: $n^{\mu}=(1,\bm{n})$ and $\tilde{n}^{\mu}=(1,-\bm{n})/2$ (note that $\phi=(nx)$). Assuming obvious differential properties of the four-vector potential $A^{\mu}(\phi)$ and its derivatives, it is clear that it is a solution of the free wave equation $\square A^{\mu}=0$, where $\square=\partial_{\nu}\partial^{\nu}$, and it is assumed to fulfill the Lorenz-gauge condition $\partial_{\mu}A^{\mu}=0$, with the additional constraint $A^0(\phi)=0$. Thus, if we represent $A^{\mu}(\phi)$ in the form $A^{\mu}(\phi)=(0,\bm{A}(\phi))$, then the Lorenz-gauge condition implies $\bm{n}\cdot\bm{A}'(\phi)=0$, with the prime in a function of $\phi$ indicating its derivative with respect to $\phi$. If we make the additional assumption that $\bm{A}(\phi)$ vanishes for $\phi\to\pm\infty$, the equality $\bm{n}\cdot\bm{A}'(\phi)=0$ implies that $\bm{n}\cdot\bm{A}(\phi)=0$. By introducing two four-vectors $a_j^{\mu}=(0,\bm{a}_j)$, with $j=1,2$, such that $(na_j)=-\bm{n}\cdot\bm{a}_j=0$ and $(a_ia_j)=-\bm{a}_i\cdot\bm{a}_j=-\delta_{ij}$, the most general form of the vector potential $\bm{A}(\phi)$ reads $\bm{A}(\phi)=\psi_1(\phi)\bm{a}_1+\psi_2(\phi)\bm{a}_2$, where the two functions $\psi_j(\phi)$ are arbitrary provided that they vanish for $\phi\to\pm\infty$ and they feature the differential properties mentioned above when the four-vector potential $A^{\mu}(\phi)$ was introduced. The field tensor $F^{\mu\nu}(\phi)=\partial^{\mu}A^{\nu}(\phi)-\partial^{\nu}A^{\mu}(\phi)$ of the plane wave is given by $F^{\mu\nu}(\phi)=n^{\mu}A^{\prime\,\nu}(\phi)-n^{\nu}A^{\prime\,\mu}(\phi)$ and below we will also use its integral $\mathscr{F}^{\mu\nu}(\phi)=\int_{-\infty}^{\phi}d\phi'F^{\mu\nu}(\phi')=n^{\mu}A^{\nu}(\phi)-n^{\nu}A^{\mu}(\phi)$ (note that the tensor $\mathscr{F}^{\mu\nu}(\phi)$ is gauge invariant).

The four-dimensional quantities $n^{\mu}$, $\tilde{n}^{\mu}$, and $a^{\mu}_j$ fulfill the completeness relation: $\eta^{\mu\nu}=n^{\mu}\tilde{n}^{\nu}+\tilde{n}^{\mu}n^{\nu}-a_1^{\mu}a_1^{\nu}-a_2^{\mu}a_2^{\nu}$ (note that $(n\tilde{n})=1$ and $(\tilde{n}a_j)=0$). Below, we will refer to the longitudinal ($n$) direction as the direction along $\bm{n}$ and to the transverse ($\perp$) plane as the plane spanned by the two perpendicular unit vectors $\bm{a}_j$. In this respect, together with the light-cone time $\phi=t-x_n$, with $x_n=\bm{n}\cdot \bm{x}$, we also introduce the remaining three light-cone coordinates $T=(\tilde{n}x)=(t+x_n)/2$, and $\bm{x}_{\perp}=(x_{\perp,1},x_{\perp,2})=-((xa_1),(xa_2))=(\bm{x}\cdot\bm{a}_1,\bm{x}\cdot\bm{a}_2)$. Analogously, the light-cone coordinates of an arbitrary four-vector $v^{\mu}=(v^0,\bm{v})$ will be indicated as $v_-=(nv)=v^0-v_n$, with $v_n=\bm{n}\cdot \bm{v}$, $v_+=(\tilde{n}v)=(v^0+v_n)/2$, and $\bm{v}_{\perp}=(v_{\perp,1},v_{\perp,2})=-((va_1),(va_2))=(\bm{v}\cdot\bm{a}_1,\bm{v}\cdot\bm{a}_2)$. Since we will employ the operator technique, it is convenient to also introduce the momenta operators $P_{\phi}=-i\partial_{\phi}=-(\tilde{n}P)=-(i\partial_t-i\partial_{x_n})/2$, $P_T=-i\partial_T=-(nP)=-(i\partial_t+i\partial_{x_n})$, and $\bm{P}_{\perp}=(P_{\perp,1},P_{\perp,2})=-i(\bm{a}_1\cdot\bm{\nabla},\bm{a}_2\cdot\bm{\nabla})$. These operators are the momenta conjugated to the light-cone coordinates in the sense that the commutator between the operator corresponding to each light-cone coordinate and the associated momentum operator is equal to the imaginary unit (all other possible commutators vanish): $[\phi,P_{\phi}]=[T,P_T]=i$ and $[X_{\perp,j},P_{\perp,k}]=i\delta_{jk}$, which are equivalent to the commutation relations $[X^{\mu},P^{\nu}]=-i\eta^{\mu\nu}$, with $P^{\mu}=i\partial^{\mu}$. 

The commutation relations $[X^{\mu},P^{\nu}]=-i\eta^{\mu\nu}$ imply that $[P^{\mu},f(X)]=i\partial_X^{\mu}f(X)$, where $f(X)$ is an arbitrary function of the four-position operator that can be expanded in Taylor series and $\partial_X^{\mu}=\partial/\partial X_{\mu}$. Analogously, it can easily be shown that $\exp[if(X)]P^{\mu}\exp[-if(X)]=P^{\mu}+\partial^{\mu}f(X)$ and then formally that $\exp[if(X)]g(P)\exp[-if(X)]=g(P+\partial f(X))$, where $g(P)$ is a function of the four-momentum that can be expanded in Taylor series [this identity has to be intended to apply to the Taylor series expansion of the function $g(P)$]. The same commutation relations imply that $\exp[ig(P)]X^{\mu}\exp[-ig(P)]=X^{\mu}-\partial_P^{\mu}g(P)$ and that $\exp[ig(P)]f(X)\exp[-ig(P)]=f(X-\partial_Pg(P))$, where $\partial_P^{\mu}=\partial/\partial P_{\mu}$ [as above, this identity has to be intended to apply to the Taylor series expansion of the function $f(X)$]. In particular, we will consider the case where the functions in the exponents are linear either in $X^{\mu}$ or in $P^{\mu}$:
\begin{align}
\label{Trans_X}
\exp(i(Xq))g(P)\exp(-i(Xq))&=g(P+q),\\
\label{Trans_P}
\exp(i(Py))f(X)\exp(-i(Py))&=f(X-y),
\end{align}
where $q^{\mu}$ and $y^{\mu}$ are constant four-vectors.

In addition, the commutation relations $[\phi,P_{\phi}]=[T,P_T]=i$ imply in particular the identities
\begin{align}
\label{Trans_phi}
\exp(ia\phi)\tilde{g}(P_{\phi})\exp(-ia\phi)&=\tilde{g}(P_{\phi}-a),\\
\label{Trans_PT}
\exp(ibP_T)\tilde{f}(T)\exp(-ibP_T)&=\tilde{f}(T+b),
\end{align}
with $a$ and $b$ being two constants and $\tilde{f}(T)$ and $\tilde{g}(P_{\phi})$ being two arbitrary functions, which we will use below.

Note that if $|x\rangle$ ($|p\rangle$) is the eigenstate of the four-position (four-momentum) operator $X^{\mu}$ ($P^{\mu}=i\partial^{\mu}$) with eigenvalue $x^{\mu}$ ($p^{\mu}$), i.e., $X^{\mu}|x\rangle=x^{\mu}|x\rangle$ ($P^{\mu}|p\rangle=p^{\mu}|p\rangle$), then, by normalizing the eigenstates $|x\rangle$ ($|p\rangle$) such that $\langle x|y\rangle=\delta^{(4)}(x-y)$ [$\langle p|q\rangle=(2\pi)^4\delta^{(4)}(p-q)$], it is $\langle x|p\rangle=\exp(-i(px))=\exp[-i(p_+\phi+p_-T-\bm{p}_{\perp}\cdot\bm{x}_{\perp})]$ and $P_{\phi}|p\rangle=-p_+|p\rangle$, $P_T|p\rangle=-p_-|p\rangle$, and $\bm{P}_{\perp}|p\rangle=\bm{p}_{\perp}|p\rangle$. Also, the operator completeness relations hold
\begin{align}
\label{C_x}
\int d^4x\,|x\rangle\langle x|&=1,\\
\int \frac{d^4p}{(2\pi)^4}\,|p\rangle\langle p|&=1.
\end{align}

The Volkov states are the exact, analytical solutions of the Dirac equation in a plane wave \cite{Volkov_1935,Landau_b_4_1982}. The positive-energy Volkov states $U_s(p,x)$ can be classified by means of the asymptotic momentum quantum numbers $\bm{p}$ (and then the energy $\varepsilon =\sqrt{m^2+\bm{p}^2}$) and of the asymptotic spin quantum number $s=1,2$ in the remote past, i.e. for $t\to-\infty$ (for notational simplicity, we have indicated the functional dependence on the four components of the electron four-momentum $p^{\mu}=(\varepsilon ,\bm{p})$, although the energy is a function of the linear momentum). Following the general notation in Ref.~\cite{Landau_b_4_1982}, these states can be written as $U_s(p,x)=E(p,x)u_s(p)$, where
\begin{equation}
\label{E_p}
E(p,x)=\bigg[1+\frac{e\hat{n}\hat{A}(\phi)}{2p_-}\bigg]\text{e}^{i\left\{-(px)-\int_{-\infty}^{\phi}d\varphi\left[\frac{e(pA(\varphi))}{p_-}-\frac{e^2A^2(\varphi)}{2p_-}\right]\right\}},
\end{equation}
and where $u_s(p)$ are the free, positive-energy spinors normalized as $u^{\dag}_s(p)u_{s'}(p)=2\varepsilon \delta_{ss'}$ \cite{Landau_b_4_1982}. In Eq. (\ref{E_p}) we have introduced the notation $\hat{v}=\gamma^{\mu}v_{\mu}$ for a generic four-vector $v^{\mu}$, with $\gamma^{\mu}$ being the Dirac matrices, which satisfy the anti-commutation relations $\{\gamma^{\mu},\gamma^{\nu}\}=2\eta^{\mu\nu}$ \cite{Landau_b_4_1982}.

The electron Green's function $G(x,x')$ in the general plane-wave background electromagnetic field described by the four-vector potential $A^{\mu}(\phi)$ is defined by the equation
\begin{equation}
\{\gamma^{\mu}[i\partial_{\mu}-eA_{\mu}(\phi)]-m\}G(x,x')=\delta^{(4)}(x-x').
\end{equation}
In order to uniquely identify the Green's function, boundary conditions have also to be specified. Here, we always assume the Feynman prescription corresponding to the shift $m\to m-i0$ \cite{Landau_b_4_1982}. Within the operator technique the operator $G$ corresponding to the Green's function $G(x,x')$ is defined via the equation $G(x,x')=\langle x|G|x'\rangle$, i.e., as
\begin{equation}
G=\frac{1}{\hat{\Pi}-m+i0},
\end{equation}
where $\Pi^{\mu}=P^{\mu}-eA^{\mu}(\Phi)$. Now, we have explicitly shown in Ref.  \cite{Di_Piazza_2018_d} (see also Refs.~\cite{Baier_1976_a,Baier_1976_b,Di_Piazza_2007}) that the operator $G$ can be written in the form
\begin{equation}
\label{G_1}
\begin{split}
G&=(\hat{\Pi}+m)\frac{1}{\hat{\Pi}^2-m^2+i0}=(\hat{\Pi}+m)(-i)\int_0^{\infty}ds\, e^{-im^2s}e^{2isP_TP_{\phi}}\\
&\times e^{-i\int_0^sds'[\bm{P}_{\perp}-e\bm{A}_{\perp}(\Phi-2s'P_T)]^2}\Big\{1-\frac{e}{2P_T}\hat{n}[\hat{A}(\Phi-2sP_T)-\hat{A}(\Phi)]\Big\},
\end{split}
\end{equation}
where the prescription $m^2\to m^2-i0$ is understood. Below, we will also need the equivalent expression
\begin{equation}
\label{G_2}
\begin{split}
G=&\frac{1}{\hat{\Pi}^2-m^2+i0}(\hat{\Pi}+m)=(-i)\int_0^{\infty}ds\, e^{-im^2s}\Big\{1+\frac{e}{2P_T}\hat{n}[\hat{A}(\Phi+2sP_T)-\hat{A}(\Phi)]\Big\}\\
&\times e^{-i\int_0^sds'[\bm{P}_{\perp}-e\bm{A}_{\perp}(\Phi+2s'P_T)]^2}e^{2isP_TP_{\phi}}(\hat{\Pi}+m).
\end{split}
\end{equation}
\section{General expression of the one-loop vertex correction}
\label{VC_General}
The one-loop vertex correction corresponds to the Feynman diagram in Fig. \ref{FD_VC}, where we have implicitly assumed that the photon four-momentum $q^{\mu}$ is outgoing. Note that the two external electron lines correspond to real electrons, i.e., the four-momenta $p^{\mu}$ and $p^{\prime\,\mu}$ are on-shell ($p^2=p^{\prime\,2}=m^2$), whereas at the moment we make no assumptions about the outgoing photon, i.e., in particular, $q^2\neq 0$. If we denote by $s$ ($s'$) the spin quantum number of the incoming (outgoing) electron and by $l$ the polarization quantum number of the outgoing photon, the amplitude $-ie\Gamma_{s,s',l}(p,p',q)$ corresponding to the Feynman diagram in Fig. \ref{FD_VC} can be written as
\begin{equation}
-ie\Gamma_{s,s',l}(p,p',q)=-e^3\int d^4x\, d^4y\, d^4z\,\bar{U}_{s'}(p',y)\gamma^{\lambda}G(y,z)\hat{e}^*_l(q)e^{i(qz)}G(z,x)\gamma^{\nu}U_s(p,x)D_{\lambda\nu}(x-y),
\end{equation}
where $e^{\mu}_l(q)$ is the polarization four-vector of the outgoing photon. Here, we have introduced the photon propagator $D^{\lambda\nu}(x)$ and we work in the Lorenz gauge such that
\begin{equation}
D^{\lambda\nu}(x)=\int\frac{d^4k}{(2\pi)^4}\frac{\eta^{\lambda\nu}}{k^2-\kappa^2+i0}e^{-i(kx)},
\end{equation}
where $\kappa^2$ is the square of a fictitious photon mass, which has been introduced to avoid infrared divergences.

By using the completeness relation in Eq. (\ref{C_x}) and the translation properties in Eq. (\ref{Trans_X}), the amplitude can be written in the semi-operator form as
\begin{equation}
\label{Gamma_2}
\begin{split}
-ie\Gamma_{s,s',l}(p,p',q)&=-e^3\int d^4x\int\frac{d^4k}{(2\pi)^4}\,\frac{1}{k^2-\kappa^2+i0}\bar{U}_{s'}(p',x)e^{i(kx)}\gamma^{\lambda}G e^{i(qx)}\hat{e}^*_l(q)Ge^{-i(kx)}\gamma_{\lambda}U_s(p,x)\\
&=-e^3\int d^4x\int\frac{d^4k}{(2\pi)^4}\,\frac{1}{k^2-\kappa^2+i0}\\
&\quad\times\bar{U}_{s'}(p',x)\gamma^{\lambda}\frac{1}{\hat{\Pi}(\phi)+\hat{k}-m+i0} e^{i(qx)}\hat{e}^*_l(q)\frac{1}{\hat{\Pi}(\phi)+\hat{k}-m+i0}\gamma_{\lambda}U_s(p,x)\\
&=-e^3\int d^4x\int\frac{d^4k}{(2\pi)^4}\,\frac{1}{k^2-\kappa^2+i0}\\
&\quad\times\bar{U}_{s'}(p',x)\gamma^{\lambda}[\hat{\Pi}(\phi)+\hat{k}+m]\frac{1}{[\hat{\Pi}(\phi)+\hat{k}]^2-m^2+i0} e^{i(qx)}\hat{e}^*_l(q)\\
&\qquad\times\frac{1}{[\hat{\Pi}(\phi)+\hat{k}]^2-m^2+i0}[\hat{\Pi}(\phi)+\hat{k}+m]\gamma_{\lambda}U_s(p,x),
\end{split}
\end{equation}
where $\Pi^{\mu}(\phi)=i\partial^{\mu}-eA^{\mu}(\phi)$. By using the fact that $[\hat{\Pi}(\phi)-m]U_s(p,x)=[\hat{\Pi}(\phi)-m]U_{s'}(p',x)=0$, we obtain
\begin{equation}
\begin{split}
-ie\Gamma_{s,s',l}(p,p',q)&=-e^3\int d^4x\int\frac{d^4k}{(2\pi)^4}\,\frac{1}{k^2-\kappa^2+i0}\\
&\quad\times\bar{U}_{s'}(p',x)[2\Pi^{\lambda}(\phi)+\gamma^{\lambda}\hat{k}]\frac{1}{[\hat{\Pi}(\phi)+\hat{k}]^2-m^2+i0} e^{i(qx)}\hat{e}^*_l(q)\\
&\qquad\times\frac{1}{[\hat{\Pi}(\phi)+\hat{k}]^2-m^2+i0}[2\Pi_{\lambda}(\phi)+\hat{k}\gamma_{\lambda}]U_s(p,x).
\end{split}
\end{equation}
Now, we notice that [see Eq. (\ref{E_p})]
\begin{equation}
\Pi^{\lambda}(\phi)U_s(p,x)=\left[\pi^{\lambda}_p(\phi)+i\frac{e\hat{n}\hat{A}'(\phi)}{2p_-}n^{\lambda}\right]U_s(p,x),
\end{equation}
where
\begin{equation}
\pi^{\lambda}_p(\phi)=p^{\lambda}-eA^{\lambda}(\phi)+\frac{e(pA(\phi))}{p_-}n^{\lambda}-\frac{e^2A^2(\phi)}{2p_-}n^{\lambda}
\end{equation}
is the classical kinetic four-momentum of an electron in the plane wave $A^{\mu}(\phi)$, with $\lim_{\phi\to\pm\infty}\pi^{\lambda}_p(\phi)=p^{\lambda}$. The kinetic four-momentum $\pi^{\lambda}_p(\phi)$ is clearly a gauge-invariant four-vector and, by using the tensor $\mathscr{F}^{\mu\nu}(\phi)$ (see Sec. \ref{Notation}), it can be written in the manifestly gauge-invariant form as
\begin{equation}
\label{pi}
\pi^{\lambda}_p(\phi)=p^{\lambda}-\frac{ep_{\mu}\mathscr{F}^{\mu\lambda}(\phi)}{p_-}+\frac{e^2p_{\mu}\mathscr{F}^{\mu\rho}(\phi)\mathscr{F}_{\rho\nu}(\phi)p^{\nu}}{2p^3_-}n^{\lambda}.
\end{equation}
In this way, the quantity $-ie\Gamma_{s,s',l}(p,p',q)$ can be written as
\begin{equation}
\begin{split}
-ie\Gamma_{s,s',l}(p,p',q)&=-e^3\int d^4x\int\frac{d^4k}{(2\pi)^4}\,\frac{1}{k^2-\kappa^2+i0}\\
&\quad\times\bar{U}_{s'}(p',x)\left[2\pi_{p'}^{\lambda}(\phi)+i\frac{e\hat{n}\hat{A}'(\phi)}{p'_-}n^{\lambda}+\gamma^{\lambda}\hat{k}\right]\frac{1}{[\hat{\Pi}(\phi)+\hat{k}]^2-m^2+i0} e^{i(qx)}\hat{e}^*_l(q)\\
&\qquad\times\frac{1}{[\hat{\Pi}(\phi)+\hat{k}]^2-m^2+i0}\left[2\pi_{p,\lambda}(\phi)+i\frac{e\hat{n}\hat{A}'(\phi)}{p_-}n_{\lambda}+\hat{k}\gamma_{\lambda}\right]U_s(p,x).
\end{split}
\end{equation}
At this point, it is convenient to use the representations in Eq. (\ref{G_1}) and in Eq. (\ref{G_2}) for the second and the first square Volkov propagator $1/\{[\hat{\Pi}(\phi)+\hat{k}]^2-m^2+i0\}$, respectively:
\begin{equation}
\begin{split}
&-ie\Gamma_{s,s',l}(p,p',q)=e^3\int d^4x\int\frac{d^4k}{(2\pi)^4}\int_0^{\infty}ds\int_0^{\infty}du\,\frac{e^{i(qx)}}{k^2-\kappa^2+i0}\\
&\quad\times\bar{U}_{s'}(p',x)\left[2\pi_{p'}^{\lambda}(\phi)+i\frac{e\hat{n}\hat{A}'(\phi)}{p'_-}n^{\lambda}+\gamma^{\lambda}\hat{k}\right]e^{-im^2s}e^{-2is(p_--q_-+k_-)(P_{\phi}-k_++q_+)}\\
&\quad\times e^{-i\int_0^sds'[\bm{p}_{\perp}-\bm{q}_{\perp}+\bm{k}_{\perp}-e\bm{A}_{\perp}(\phi+2s'(p_--q_-+k_-))]^2}\left\{1+\frac{e\hat{n}[\hat{A}(\phi+2s(p_--q_-+k_-))-\hat{A}(\phi)]}{2(p_--q_-+k_-)}\right\}\\
&\quad\times \hat{e}^*_l(q)e^{-im^2u}\left\{1-\frac{e\hat{n}[\hat{A}(\phi-2u(p_-+k_-))-\hat{A}(\phi)]}{2(p_-+k_-)}\right\}e^{-i\int_0^udu'[\bm{p}_{\perp}+\bm{k}_{\perp}-e\bm{A}_{\perp}(\phi-2u'(p_-+k_-))]^2}\\
&\quad\times e^{-2iu(p_-+k_-)(P_{\phi}-k_+)}\left[2\pi_{p,\lambda}(\phi)+i\frac{e\hat{n}\hat{A}'(\phi)}{p_-}n_{\lambda}+\hat{k}\gamma_{\lambda}\right]U_s(p,x),
\end{split}
\end{equation}
where we have exploited the fact that Volkov states are eigenstates of the operators $P_T$ and $\bm{P}_{\perp}$. Indeed, the only operator remaining in this equation is $P_{\phi}$. Now, we use the translation property in Eq. (\ref{Trans_P}) and, analogously to the vacuum case, we write the amplitude $-ie\Gamma_{s,s',l}(p,p',q)$ in the form
\begin{equation}
\label{Gamma^mu}
-ie\Gamma_{s,s',l}(p,p',q)=-ie\int d^4x\,e^{i(qx)}\bar{U}_{s'}(p',x)\Gamma^{\mu}(p,p',q;\phi)U_s(p,x)e^*_{l,\mu}(q),
\end{equation}
where
\begin{equation}
\label{Gamma_3}
\begin{split}
-ie\Gamma^{\mu}(p,p',q;\phi)&=e^3\int\frac{d^4k}{(2\pi)^4}\int_0^{\infty}ds\int_0^{\infty}du\,\frac{e^{-im^2(s+u)}}{k^2-\kappa^2+i0}\\
&\quad\times e^{2ik_+[s(p'_-+k_-)+u(p_-+k_-)]}e^{i\left\{p'_+(\phi_s-\phi)+\int_{\phi}^{\phi_s}d\phi'\left[-\frac{e\bm{p}'_{\perp}\cdot\bm{A}_{\perp}(\phi')}{p'_-}+\frac{e^2\bm{A}^2_{\perp}(\phi')}{2p'_-}\right]\right\}}\\
&\quad\times e^{-i\int_0^sds'[\bm{p}'_{\perp}+\bm{k}_{\perp}-e\bm{A}_{\perp}(\phi_{s'})]^2}e^{-i\int_0^udu'[\bm{p}_{\perp}+\bm{k}_{\perp}-e\bm{A}_{\perp}(\phi_{u'})]^2}\\
&\quad\times e^{i\left\{-p_+(\phi_u-\phi)-\int_{\phi}^{\phi_u}d\phi'\left[-\frac{e\bm{p}_{\perp}\cdot\bm{A}_{\perp}(\phi')}{p_-}+\frac{e^2\bm{A}^2_{\perp}(\phi')}{2p_-}\right]\right\}}M^{\mu}(\phi,k,s,u).
\end{split}
\end{equation}
Here, we have introduced the quantities
\begin{align}
\label{phi_s}
\phi_s&=\phi+2s(p'_-+k_-),\\
\label{phi_u}
\phi_u&=\phi-2u(p_-+k_-),
\end{align}
and the matrix
\begin{equation}
\begin{split}
M^{\mu}(k,s,u;\phi)&=\left\{1-\frac{e\hat{n}[\hat{A}(\phi_s)-\hat{A}(\phi)]}{2p'_-}\right\}\left[2\pi_{p'}^{\lambda}(\phi_s)+i\frac{e\hat{n}\hat{A}'(\phi_s)}{p'_-}n^{\lambda}+\gamma^{\lambda}\hat{k}\right]\\
&\quad\times\left\{1+\frac{e\hat{n}[\hat{A}(\phi_s)-\hat{A}(\phi)]}{2(p'_-+k_-)}\right\}\gamma^{\mu}\left\{1-\frac{e\hat{n}[\hat{A}(\phi_u)-\hat{A}(\phi)]}{2(p_-+k_-)}\right\}\\
&\quad\times \left[2\pi_{p,\lambda}(\phi_u)+i\frac{e\hat{n}\hat{A}'(\phi_u)}{p_-}n_{\lambda}+\hat{k}\gamma_{\lambda}\right]\left\{1+\frac{e\hat{n}[\hat{A}(\phi_u)-\hat{A}(\phi)]}{2p_-}\right\}.
\end{split}
\end{equation}
Note that the integrals in $T$ and $\bm{x}_{\perp}$ can be easily taken and enforce the conservation laws $p_-=p'_-+q_-$ and $\bm{p}_{\perp}=\bm{p}'_{\perp}+\bm{q}_{\perp}$, typical of problems in a plane-wave background field. As a related remark, it is clear that the quantity $\Gamma^{\mu}(p,p',q;\phi)$, unlike the corresponding vacuum expression, depends also on the plane-wave phase. Finally, the definition in Eq. (\ref{Gamma^mu}) is consistent with the idea that computing the amplitude of the vertex in a plane wave up to one loop, one can use the substitution rule $\gamma^{\mu}\to \gamma^{\mu}+\Gamma^{\mu}(p,p',q;\phi)$, as for the vertex correction in vacuum.

The phase in Eq. (\ref{Gamma_3}) can be written in a compact form by turning the integral from $\phi$ to $\phi_s$ (from $\phi$ to $\phi_u$) into an integral in $s'$ ($u'$) like that in the third line of Eq. (\ref{Gamma_3}). By exponentiating also the denominator $k^2-\kappa^2+i0$ in the photon propagator, the quantity $\Gamma^{\mu}(p,p',q;\phi)$ can be written as
\begin{equation}
\Gamma^{\mu}(p,p',q;\phi)=e^2\int\frac{d^4k}{(2\pi)^4}\int_0^{\infty}ds\int_0^{\infty}du\int_0^{\infty}dt\,e^{iSk^2-i\kappa^2t+2i(k\tilde{F})} M^{\mu}(k,s,u;\phi),
\end{equation}
where $S=u+s+t$ and
\begin{equation}
\label{F}
\tilde{F}^{\mu}=\int_0^sds'\pi_{p'}^{\mu}(\phi_{s'})+\int_0^udu'\pi_p^{\mu}(\phi_{u'}).
\end{equation}
As next step, we can perform the integrals in $d^4k$ analytically by shifting the four-momentum $k^{\mu}$ by setting $k^{\prime\mu}=k^{\mu}+\tilde{F}^{\mu}/S$, which, since all components of $\tilde{F}^{\mu}$ except $\tilde{F}_-$ depend on $k_-$, implies that $k^{\mu}=k^{\prime\mu}-\tilde{G}^{\mu}/S$, where
\begin{equation}
\label{tG}
\tilde{G}^{\mu}=\int_0^sds'\pi_{p'}^{\mu}(\tilde{\psi}_{s'})+\int_0^udu'\pi_p^{\mu}(\tilde{\psi}_{u'}),
\end{equation}
such that $\tilde{G}_-=\tilde{F}_-=sp'_-+up_-$. Here, we have introduced the two shifted phases
\begin{align}
\label{tpsi_p}
\tilde{\psi}_s&=\phi+2s\tau'_-+2sk_-,\\
\label{tpsi_pp}
\tilde{\psi}_u&=\phi-2u\tau_--2uk_-,
\end{align}
where
\begin{align}
\label{tau_p_m}
\tau'_-&=p'_--\frac{\tilde{G}_-}{S}=\frac{tp'_--uq_-}{S},
\\
\label{tau_m}
\tau_-&=p_--\frac{\tilde{G}_-}{S}=\frac{tp_-+sq_-}{S}.
\end{align}
After the shift of the four-momentum $k^{\mu}$, we can write $\Gamma^{\mu}(p,p',q;\phi)$ in the form
\begin{equation}
\Gamma^{\mu}(p,p',q;\phi)=e^2\int_0^{\infty}dsdudt\int\frac{d^4k}{(2\pi)^4}e^{-i\kappa^2t-i\frac{\tilde{G}^2}{S}+iSk^2}\tilde{L}(\tilde{Q}^{\prime\lambda}+\gamma^{\lambda}\hat{k})\tilde{C}^{\mu}(\tilde{Q}_{\lambda}+\hat{k}\gamma_{\lambda})\tilde{R},
\end{equation}
where
\begin{align}
\label{tL}
\tilde{L}&=1-\frac{e\hat{n}[\hat{A}(\tilde{\psi}_s)-\hat{A}(\phi)]}{2p'_-},\\
\tilde{Q}^{\lambda}&=2\pi_p^{\lambda}(\tilde{\psi}_u)+i\frac{e\hat{n}\hat{A}'(\tilde{\psi}_u)}{p_-}n^{\lambda}-\frac{\hat{\tilde{G}}}{S}\gamma^{\lambda},\\
\tilde{C}^{\mu}&=\left\{1+\frac{e\hat{n}[\hat{A}(\tilde{\psi}_s)-\hat{A}(\phi)]}{2(\tau'_-+k_-)}\right\}\gamma^{\mu}\left\{1-\frac{e\hat{n}[\hat{A}(\tilde{\psi}_u)-\hat{A}(\phi)]}{2(\tau_-+k_-)}\right\},\\
\tilde{Q}^{\prime\lambda}&=2\pi_{p'}^{\lambda}(\tilde{\psi}_s)+i\frac{e\hat{n}\hat{A}'(\tilde{\psi}_s)}{p'_-}n^{\lambda}-\gamma^{\lambda}\frac{\hat{\tilde{G}}}{S},\\
\label{tR}
\tilde{R}&=1+\frac{e\hat{n}[\hat{A}(\tilde{\psi}_u)-\hat{A}(\phi)]}{2p_-}.
\end{align}

The integral in $d^4k$ in $\Gamma^{\mu}(p,p',q;\phi)$ is complicated by the fact that the variable $k_-$ is contained in the argument of the four-vector potential of the plane wave. Thus, we first take the integral in $d^2\bm{k}_{\perp}$, which is Gaussian:
\begin{equation}
\Gamma^{\mu}(p,p',q;\phi)=-i\alpha\int_0^{\infty}\frac{dsdudt}{S}\int\frac{dk_-dk_+}{(2\pi)^2}e^{-i\kappa^2t-i\frac{\tilde{G}^2}{S}+2iSk_-k_+}\tilde{M}^{\mu}(k_-,k_+,s,u,t;\phi),
\end{equation}
where $\alpha=e^2/4\pi\approx 1/137$ is the fine-structure constant and where
\begin{equation}
\begin{split}
\tilde{M}^{\mu}(k_-,k_+,s,u,t;\phi)=&\tilde{L}\left[(\tilde{Q}^{\prime\lambda}+k_-\gamma^{\lambda}\hat{\tilde{n}})\tilde{C}^{\mu}(\tilde{Q}_{\lambda}+k_-\hat{\tilde{n}}\gamma_{\lambda})-\frac{i}{2S}\gamma^{\lambda}\gamma_{\perp,i}\tilde{C}^{\mu}\gamma_{\perp,i}\gamma_{\lambda}\right]\tilde{R}\\
&+k_+\tilde{L}[\gamma^{\lambda}\hat{n}\tilde{C}^{\mu}(\tilde{Q}_{\lambda}+k_-\hat{\tilde{n}}\gamma_{\lambda})+(\tilde{Q}^{\prime\lambda}+k_-\gamma^{\lambda}\hat{\tilde{n}})\tilde{C}^{\mu}\hat{n}\gamma_{\lambda}]\tilde{R}\\
&+k_+^2\tilde{L}\gamma^{\lambda}\hat{n}\gamma^{\mu}\hat{n}\gamma_{\lambda}\tilde{R}.
\end{split}
\end{equation}
Finally, the integral in $dk_+$ results in a delta function and its first and second derivatives all evaluated at $2Sk_-$. This allows then also to take the integral in $dk_-$ and, after straightforward manipulations, the resulting expression of $\Gamma^{\mu}(p,p',q;\phi)$ can be written as
\begin{equation}
\label{Gamma_f}
\begin{split}
\Gamma^{\mu}(p,p',q;\phi)&=-\frac{i\alpha}{4\pi}\int_0^{\infty}\frac{dsdudt}{S^3}\,e^{-i\kappa^2t}\bigg\{e^{-i\frac{\tilde{G}^2}{S}}\tilde{L}(S\tilde{Q}^{\prime\lambda}\tilde{C}^{\mu}\tilde{Q}_{\lambda}+2i\tilde{C}^{\mu})\tilde{R}\\
&\quad\left.\left.+\frac{i}{2}\frac{d}{dk_-}\left[e^{-i\frac{\tilde{G}^2}{S}}\tilde{L}(\gamma^{\lambda}\hat{n}\tilde{C}^{\mu}\tilde{Q}_{\lambda}+\tilde{Q}^{\prime\lambda}\tilde{C}^{\mu}\hat{n}\gamma_{\lambda})\tilde{R}\right]+\frac{\hat{n}}{S}n^{\mu}\frac{d^2}{dk^2_-}\left(e^{-i\frac{\tilde{G}^2}{S}}\right)\right\}\right\vert_{k_-=0}\\
&=-\frac{i\alpha}{4\pi}\int_0^{\infty}\frac{dsdudt}{S^3}\,e^{-i\kappa^2t-i\frac{\tilde{G}^2}{S}}\\
&\quad\times\Bigg\{\tilde{L}\left[S\tilde{Q}^{\prime\lambda}\tilde{C}^{\mu}\tilde{Q}_{\lambda}+2i\tilde{C}^{\mu}+\frac{1}{2S}\frac{d\tilde{G}^2}{dk_-}(\gamma^{\lambda}\hat{n}\tilde{C}^{\mu}\tilde{Q}_{\lambda}+\tilde{Q}^{\prime\lambda}\tilde{C}^{\mu}\hat{n}\gamma_{\lambda})\right]\tilde{R}\\
&\quad\left.\left.+\frac{i}{2}\frac{d}{dk_-}\left[\tilde{L}(\gamma^{\lambda}\hat{n}\tilde{C}^{\mu}\tilde{Q}_{\lambda}+\tilde{Q}^{\prime\lambda}\tilde{C}^{\mu}\hat{n}\gamma_{\lambda})\tilde{R}\right]-\frac{\hat{n}}{S^2}n^{\mu}\left[\frac{1}{S}\left(\frac{d\tilde{G}^2}{dk_-}\right)^2+i\frac{d^2\tilde{G}^2}{dk^2_-}\right]\right\}\right\vert_{k_-=0}.
\end{split}
\end{equation}
This expression can be further manipulated especially to simplify its matrix structure. However, it is first convenient to make the following considerations related to the Ward identity to be fulfilled by $\Gamma^{\mu}(p,p',q;\phi)$ \cite{Itzykson_b_1980}. From now on we assume that $q_->0$. Thus, by using the three four-vectors
\begin{align}
N^{\mu}&=q^{\mu}-\frac{q^2n^{\mu}}{2q_-},\\
\label{Lambda_i}
\Lambda^{\mu}_i&=a_i^{\mu}+\frac{q_{\perp,i}n^{\mu}}{q_-},
\end{align}
with $i=1,2$ together with $n^{\mu}$, one can build a light-cone basis such that
\begin{equation}
\eta^{\mu\nu}=\frac{N^{\mu}n^{\nu}+n^{\mu}N^{\nu}}{q_-}-\Lambda^{\mu}_1\Lambda^{\nu}_1-\Lambda^{\mu}_2\Lambda^{\nu}_2.
\end{equation}
Then, the quantity $\Gamma^{\mu}(p,p',q;\phi)e^*_{l,\mu}(q)$, which is the one finally required here, can be written as [recall that we work in the Lorenz gauge where $(qe^*_l(q))=0$]
\begin{equation}
\label{Gamma_exp}
\begin{split}
\Gamma^{\mu}(p,p',q;\phi)e^*_{l,\mu}(q)&=\frac{e^*_{l,-}(q)}{q_-}\Gamma_q(p,p',q;\phi)-\frac{q^2e^*_{l,-}(q)}{q_-^2}\Gamma_-(p,p',q;\phi)\\
&\quad-(\Gamma(p,p',q;\phi)\Lambda_1)(\Lambda_1e^*_l(q))-(\Gamma(p,p',q;\phi)\Lambda_2)(\Lambda_2e^*_l(q))\\
&=\frac{e^*_{l,-}(q)}{q_-}\Gamma_q(p,p',q;\phi)\\
&\quad-\left[\frac{q^2e^*_{l,-}(q)}{q_-^2}+\frac{q_{\perp,1}}{q_-}(\Lambda_1e^*_l(q))+\frac{q_{\perp,2}}{q_-}(\Lambda_2e^*_l(q))\right]\Gamma_-(p,p',q;\phi)\\
&\quad+\Gamma_{\perp,1}(p,p',q;\phi)(\Lambda_1e^*_l(q))+\Gamma_{\perp,2}(p,p',q;\phi)(\Lambda_2e^*_l(q))
\end{split}
\end{equation}
where $\Gamma_q(p,p',q;\phi)=(\Gamma^{\mu}(p,p',q;\phi)q_{\mu})$ and all other symbols are defined in analogy to the definitions given in the introduction. Now, since the structure of the function $\Gamma^{\mu}(p,p',q;\phi)$ is complicated because of the presence of the plane wave, it is clear that the components $\Gamma_-(p,p',q;\phi)=(\Gamma^{\mu}(p,p',q;\phi)n_{\mu})$ and $\Gamma_{\perp,j}(p,p',q;\phi)=-(\Gamma^{\mu}(p,p',q;\phi)a_{j,\mu})$ are relatively easy to work out because the quantities $n^{\mu}$ and $a_i^{\mu}$ characterize the plane wave. For example, we observe that all the terms proportional to $n^{\mu}$ in $\Gamma^{\mu}(p,p',q;\phi)$ can be ignored in the computation of the components $\Gamma_-(p,p',q;\phi)$ and $\Gamma_{\perp,j}(p,p',q;\phi)$. 

The apparently most complicated term is, therefore, $\Gamma_q(p,p',q;\phi)$, which is related to the fact that by itself the vertex-correction function is not gauge invariant. However, the gauge invariance of QED guarantees that the component $\Gamma_q(p,p',q;\phi)$ of the vertex-correction function does not to contribute to any transition amplitude involving on-shell external electrons/positrons. We explicitly prove this statement in the case under consideration with an incoming and an outgoing electron, the other possible cases being proved in an analogous way. First, we start back from Eq. (\ref{Gamma_2}) and we apply the same procedure to prove the Ward identity \cite{Mitter_1975,Morozov_1981}. From the second equality in Eq. (\ref{Gamma_2}) and from the definition of $\Gamma^{\mu}(p,p',q;\phi)$ in Eq. (\ref{Gamma^mu}), we obtain
\begin{equation}
\Gamma_q(p,p',q;\phi)=-ie^2\int\frac{d^4k}{(2\pi)^4}\,\frac{1}{k^2-\kappa^2+i0}\gamma^{\lambda}\frac{1}{\hat{\Pi}(\phi)+\hat{k}-\hat{q}-m+i0}\hat{q}\frac{1}{\hat{\Pi}(\phi)+\hat{k}-m+i0}\gamma_{\lambda}.
\end{equation}
now, by writing $\hat{q}=\hat{\Pi}(\phi)+\hat{k}-m-[\hat{\Pi}(\phi)+\hat{k}-\hat{q}-m]$ it is clear that we can express $\Gamma_q(p,p',q;\phi)$ as the difference of two terms containing only one propagator in the plane wave:
\begin{equation}
\label{Gamma_q_0}
\Gamma_q(p,p',q;\phi)=-ie^2\int\frac{d^4k}{(2\pi)^4}\,\frac{1}{k^2-\kappa^2+i0}\gamma^{\lambda}\left[\frac{1}{\hat{\Pi}(\phi)+\hat{k}-\hat{q}-m+i0}-\frac{1}{\hat{\Pi}(\phi)+\hat{k}-m+i0}\right]\gamma_{\lambda}.
\end{equation}

\begin{figure}
\begin{center}
\includegraphics[width=0.4\columnwidth]{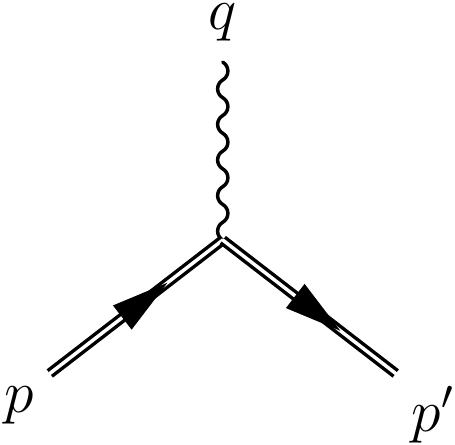}
\caption{The leading-order Feynman diagram corresponding to nonlinear Compton scattering of an off-shell photon. The double lines represent exact electron states in a plane wave (Volkov states) \cite{Landau_b_4_1982}.}
\label{FD_NCS_LO}
\end{center}
\end{figure}

\begin{figure}
\begin{center}
\includegraphics[width=\columnwidth]{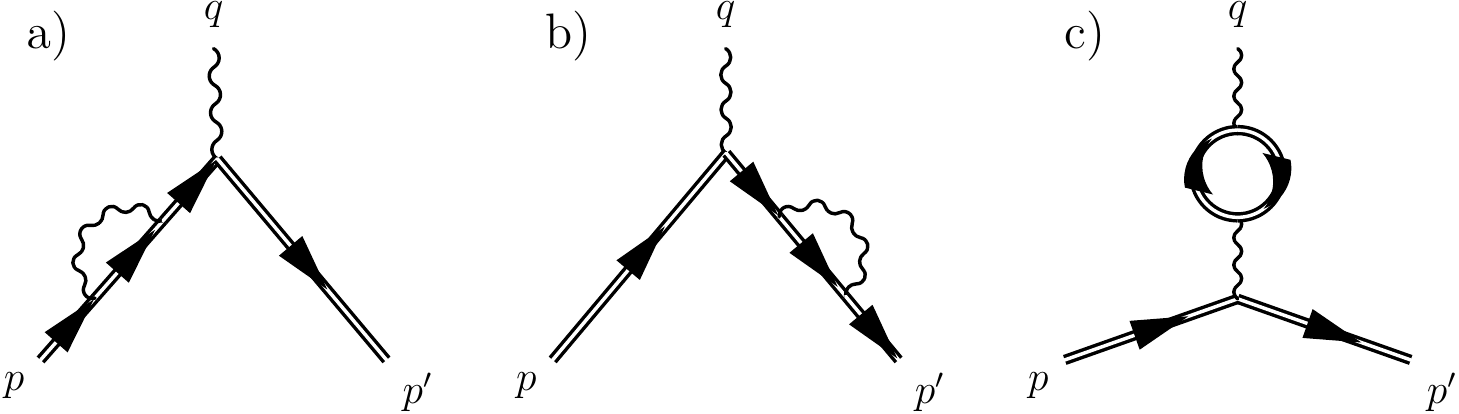}
\caption{The one-loop Feynman diagrams corresponding, together with the one-loop vertex correction in Fig. \ref{FD_VC}, to the leading-order radiative corrections of nonlinear Compton scattering of an off-shell photon. The double lines represent exact electron states and propagator in a plane wave (Volkov states and propagator, respectively) \cite{Landau_b_4_1982}.}
\label{FD_NCS_MO_PO}
\end{center}
\end{figure}

At this point, we observe that in computing, for example, the one-loop radiative corrections to nonlinear Compton scattering (see the leading-order diagram in Fig. \ref{FD_NCS_LO}, for the kinematical situation corresponding to the case under study)
we also have to include the remaining diagrams listed in Fig. \ref{FD_NCS_MO_PO}.
If we indicate as $i\mathcal{M}^{(1)}_{s,s',\mu}(p,p',q)e_l^{*\,\mu}(q)$ the amplitude of  the one-loop radiative corrections to nonlinear Compton scattering represented by the diagrams in Figs. \ref{FD_VC} and \ref{FD_NCS_MO_PO}, the gauge invariance of QED implies that $\mathcal{M}^{(1)}_{s,s',\mu}(p,p',q)q^{\mu}=0$ \cite{Peskin_b_1995}. Since the contribution corresponding to Fig. \ref{FD_NCS_MO_PO}.c is by itself gauge invariant \cite{Baier_1976_b,Meuren_2013}, by summing the contributions from Fig. \ref{FD_VC} and from Figs. \ref{FD_NCS_MO_PO}.b and \ref{FD_NCS_MO_PO}.c, we obtain [see also Eq. (\ref{Gamma_q_0})]
\begin{equation}
\begin{split}
i\mathcal{M}^{(1)}_{s,s',\mu}(p,p',q)q^{\mu}&=-e^3\int d^4x\int\frac{d^4k}{(2\pi)^4}\,\frac{1}{k^2-\kappa^2+i0}\\
&\times\bar{U}_{s'}(p',x)\left\{\gamma^{\lambda}\left[\frac{1}{\hat{\Pi}(\phi)+\hat{k}-m+i0}e^{i(qx)}-e^{i(qx)}\frac{1}{\hat{\Pi}(\phi)+\hat{k}-m+i0}\right]\gamma_{\lambda}\right.\\
&\quad+\hat{q}e^{i(qx)}\frac{1}{\hat{\Pi}(\phi)-m+i0}\gamma^{\lambda}\frac{1}{\hat{\Pi}(\phi)+\hat{k}-m+i0}\gamma_{\lambda}\\
&\quad\left.+\gamma^{\lambda}\frac{1}{\hat{\Pi}(\phi)+\hat{k}-m+i0}\gamma_{\lambda}\frac{1}{\hat{\Pi}(\phi)-m+i0}\hat{q}e^{i(qx)}\right\}U_s(p,x).
\end{split}
\end{equation}
Now, by using the fact that $[\hat{\Pi}(\phi)-m]U_s(p,x)=[\hat{\Pi}(\phi)-m]U_{s'}(p',x)=0$, we first replace $\hat{q}$ with $\hat{\Pi}(\phi)+\hat{q}-m$ ($m-\hat{\Pi}(\phi)+\hat{q}$) in the third (fourth) line of this equation and then we move the exponential $\exp[i(qx)]$ to the left of all other operators by exploiting the identity in Eq. (\ref{Trans_X}). The result is
\begin{equation}
\begin{split}
i\mathcal{M}^{(1)}_{s,s',\mu}(p,p',q)q^{\mu}&=-e^3\int d^4x\int\frac{d^4k}{(2\pi)^4}\,\frac{e^{i(qx)}}{k^2-\kappa^2+i0}\\
&\times\bar{U}_{s'}(p',x)\left\{\gamma^{\lambda}\left[\frac{1}{\hat{\Pi}(\phi)+\hat{k}-\hat{q}-m+i0}-\frac{1}{\hat{\Pi}(\phi)+\hat{k}-m+i0}\right]\gamma_{\lambda}\right.\\
&\quad\left.+\gamma^{\lambda}\frac{1}{\hat{\Pi}(\phi)+\hat{k}-m+i0}\gamma_{\lambda}-\gamma^{\lambda}\frac{1}{\hat{\Pi}(\phi)+\hat{k}-\hat{q}-m+i0}\gamma_{\lambda}\right\}U_s(p,x),
\end{split}
\end{equation}
which indeed vanishes identically. This result indicates that gauge invariance implies that the component $\Gamma_q(p,p',q;\phi)$ can be ignored as it will always be compensated by the corresponding contributions arising from the mass operators (some properties of $\Gamma_q(p,p',q;\phi)$ are discussed in the appendix).

At this point, we can consider the other component $\Gamma_-(p,p',q;\phi)$, whose structure is particularly easy. In fact, starting from Eq. (\ref{Gamma_f}), we have that
\begin{equation}
\label{Gamma_-}
\begin{split}
\Gamma_-(p,p',q;\phi)&=-\frac{i\alpha}{4\pi}\int_0^{\infty}\frac{dsdudt}{S^3}\,e^{-i\kappa^2t-i\frac{G^2}{S}}\left(SLQ^{\prime\lambda}\hat{n}Q_{\lambda}R+2i\hat{n}\right)\\
&=-\frac{i\alpha}{2\pi}\int_0^{\infty}\frac{dsdudt}{S^3}\,e^{-i\kappa^2t-i\frac{G^2}{S}}\left[\left(2S(\pi_s\pi_u)+\frac{G^2}{S}+i\right)\hat{n}-2G_-(\hat{\pi}_sR+L\hat{\pi}_u)\right.\\
&\left.\quad-\hat{G}\hat{\pi}_s\hat{n}-\hat{n}\hat{\pi}_u\hat{G}+2\tau_-L\hat{G}+2\tau'_-\hat{G}R+2\frac{G_-}{S}\hat{G}-\frac{G_-^2}{S}\frac{\hat{\Delta}_s\hat{n}\hat{\Delta}_u}{p_-p'_-}\right],
\end{split}
\end{equation}
where
\begin{align}
\pi_s^{\mu}&=\pi^{\mu}_{p'}(\psi_s),\quad \Delta^{\mu}_s=e[A^{\mu}(\psi_s)-A^{\mu}(\phi)], & L&=1-\frac{\hat{n}\hat{\Delta}_s}{2p'_-}, \quad \psi_s=\phi+2s\tau'_-,\\
\pi_u^{\mu}&=\pi^{\mu}_p(\psi_u),\quad \Delta^{\mu}_u=e[A^{\mu}(\psi_u)-A^{\mu}(\phi)], & R&=1+\frac{\hat{n}\hat{\Delta}_u}{2p_-},\quad \psi_u=\phi-2u\tau_-,\\
C^{\mu}&=\left(1+\frac{\hat{n}\hat{\Delta}_s}{2\tau'_-}\right)\gamma^{\mu}\left(1-\frac{\hat{n}\hat{\Delta}_u}{2\tau_-}\right), & G^{\mu}&=\int_0^sds'\pi^{\mu}_{p'}(\psi_{s'})+\int_0^udu'\pi^{\mu}_p(\psi_{u'}), \\
Q^{\lambda}&=2\pi_u^{\lambda}+i\frac{e\hat{n}\hat{A}'(\psi_u)}{p_-}n^{\lambda}-\frac{\hat{G}}{S}\gamma^{\lambda},&  
Q^{\prime\lambda}&=2\pi_s^{\lambda}+i\frac{e\hat{n}\hat{A}'(\psi_s)}{p'_-}n^{\lambda}-\gamma^{\lambda}\frac{\hat{G}}{S}.
\end{align}

As we have mentioned, in order to compute the components $\Gamma_{\perp,j}(p,p',q;\phi)=-(\Gamma^{\mu}(p,p',q;\phi)a_{j,\mu})$ we can effectively assume that the matrix $\hat{n}$ anticommutes with $\gamma^{\mu}$. In the following four equations, with an abuse of notation, we use the equal symbol also for two matrices that are equal to each other up to terms proportional to $n^{\mu}$, which can anyway be ignored in the computation of $\Gamma_{\perp,j}(p,p',q;\phi)$. Going through the terms in Eq. (\ref{Gamma_f}) in order of complexity, one can easily show that
\begin{align}
&LC^{\mu}R=\gamma^{\mu}+\frac{G_-}{2p'_-\tau'_-S}\hat{n}\hat{\Delta}_s\gamma^{\mu}-\frac{G_-}{2p_-\tau_-S}\gamma^{\mu}\hat{n}\hat{\Delta}_u,\\
\begin{split}
&L(\gamma^{\lambda}\hat{n}C^{\mu}Q_{\lambda}+Q^{\prime\lambda}C^{\mu}\hat{n}\gamma_{\lambda})R=-\frac{2\tau_-}{p'_-}\hat{n}\hat{\Delta}_s\gamma^{\mu}+\frac{2\tau'_-}{p_-}\gamma^{\mu}\hat{n}\hat{\Delta}_u-\frac{4G_-}{S}\gamma^{\mu}+\frac{4G^{\mu}}{S}\hat{n}\\
&\qquad-2\hat{n}\gamma^{\mu}\hat{\pi}_s-2\hat{\pi}_u\gamma^{\mu}\hat{n},
\end{split}\\
\begin{split}
&\frac{d}{dk_-}\left[\tilde{L}(\gamma^{\lambda}\hat{n}\tilde{C}^{\mu}\tilde{Q}_{\lambda}+\tilde{Q}^{\prime\lambda}\tilde{C}^{\mu}\hat{n}\gamma_{\lambda})\tilde{R}\right]_{k_-=0}=8\left(\frac{G_1^{\mu}}{S}-\frac{s\tau_-}{p'_-}\mathcal{A}_s^{\prime\,\mu}+\frac{u\tau'_-}{p_-}\mathcal{A}_u^{\prime\,\mu}\right)\hat{n}\\
&\qquad+4s\left(1+\frac{\tau_-}{p'_-}\right)\hat{n}\gamma^{\mu}\hat{\mathcal{A}}_s'-4u\left(1+\frac{\tau'_-}{p_-}\right)\hat{\mathcal{A}}_u'\gamma^{\mu}\hat{n},
\end{split}\\
\begin{split}
&LQ^{\prime\,\lambda}C^{\mu}Q_{\lambda}R=4(\pi_s\pi_u)\left(\gamma^{\mu}+\frac{G_-}{2p'_-\tau'_-S}\hat{n}\hat{\Delta}_s\gamma^{\mu}-\frac{G_-}{2p_-\tau_-S}\gamma^{\mu}\hat{n}\hat{\Delta}_u\right)\\
&\qquad+2i\frac{\tau_-}{p'_-}\hat{n}\hat{\mathcal{A}}'_s\gamma^{\mu}+2i\frac{\tau'_-}{p_-}\gamma^{\mu}\hat{n}\hat{\mathcal{A}}'_u-\frac{2}{S}LC^{\mu}\hat{G}\hat{\pi}_sR-\frac{2}{S}L\hat{\pi}_u\hat{G}C^{\mu}R\\
&\qquad-\frac{2}{S^2}L\hat{G}\left(\gamma^{\mu}+\frac{\gamma^{\mu}\hat{\Delta}_s\hat{n}}{2\tau'_-}-\frac{\hat{\Delta}_u\hat{n}\gamma^{\mu}}{2\tau_-}\right)\hat{G}R
\end{split}
\end{align}
where
\begin{equation}
\begin{split}
G_1^{\mu}&=\frac{d}{d\phi}\left[\int_0^sds'\,s'\pi^{\mu}_{p'}(\psi_{s'})-\int_0^udu'\,u'\pi^{\mu}_p(\psi_{u'})\right]\\
&=\frac{1}{2\tau'_-}\left[s\pi_{p'}^{\mu}(\psi_s)-\int_0^sds'\pi_{p'}^{\mu}(\psi_{s'})\right]+\frac{1}{2\tau_-}\left[u\pi_p^{\mu}(\psi_u)-\int_0^udu'\pi_p^{\mu}(\psi_{u'})\right]
\end{split}
\end{equation}
and
\begin{equation}
\mathcal{A}^{\mu}_{s/u}=eA^{\mu}(\psi_{s/u})
\end{equation}
(the prime on these quantities indicates the derivative with respect to $\phi$).

In this way, we obtain the following expressions of the transverse components $\Gamma_{\perp,i}(p,p',q;\phi)$:
\begin{equation}
\label{Gamma_j}
\begin{split}
\Gamma_{\perp,j}(p,p',q;\phi)&=\frac{i\alpha}{2\pi}\int_0^{\infty}\frac{dsdudt}{S^3}\,e^{-i\kappa^2t-i\frac{G^2}{S}}\\
&\quad\times\left\{(2S(\pi_s\pi_u)+i)\left(\hat{a}_j+\frac{G_-}{2p'_-\tau'_-S}\hat{n}\hat{\Delta}_s\hat{a}_j-\frac{G_-}{2p_-\tau_-S}\hat{a}_j\hat{n}\hat{\Delta}_u\right)\right.\\
&\qquad-L(Ca_j)\hat{G}\hat{\pi}_sR-L\hat{\pi}_u\hat{G}(Ca_j)R-\frac{1}{S}L\hat{G}\left(\hat{a}_j+\frac{\hat{a}_j\hat{\Delta}_s\hat{n}}{2\tau'_-}-\frac{\hat{\Delta}_u\hat{n}\hat{a}_j}{2\tau_-}\right)\hat{G}R\\
&\qquad-\frac{2(GG_1)}{S}\left(\frac{\tau_-}{p'_-}\hat{n}\hat{\Delta}_s\hat{a}_j-\frac{\tau'_-}{p_-}\hat{a}_j\hat{n}\hat{\Delta}_u+\frac{2G_-}{S}\hat{a}_j-\frac{2(Ga_j)}{S}\hat{n}+\hat{n}\hat{a}_j\hat{\pi}_s+\hat{\pi}_u\hat{a}_j\hat{n}\right)\\
&\qquad+2i\left(\frac{(G_1a_j)}{S}+s(\mathcal{A}'_sa_j)-u(\mathcal{A}'_ua_j)\right)\hat{n}\\
&\qquad\left.+i\left[s-(u+t)\frac{\tau_-}{p'_-}\right]\hat{\mathcal{A}}_s'\hat{n}\hat{a}_j-i\left[u-(s+t)\frac{\tau'_-}{p_-}\right]\hat{a}_j\hat{n}\hat{\mathcal{A}}_u'\right\}
\end{split}
\end{equation}
Equations (\ref{Gamma_exp}), (\ref{Gamma_-}), and (\ref{Gamma_j}) are the main results of the paper and, as it can easily be shown, they reduce to the result in vacuum as, e.g., on page 339 of Ref. \cite{Itzykson_b_1980} [as it is shown in the appendix, the component $\Gamma_q(p,p',q;\phi)$ vanishes in vacuum]. We notice that the pre-exponential matrices in all terms feature symmetry properties such that they can all be written as the sum of two classes of terms with the second one, being obtained from the first one by: 1) taking the Dirac conjugate, 2) swapping all indexes $s$ and $u$ in each quantity. We have exploited this symmetry in the computations presented below. Finally, we observe that all the terms in Eqs. (\ref{Gamma_exp}), (\ref{Gamma_-}), and (\ref{Gamma_j}) have at most three gamma matrices except the three terms on the third line of Eq. (\ref{Gamma_j}) \footnote{The terms with three gamma matrices can be further reduced according to the identity
\begin{equation*}
\begin{split}
\hat{A}\hat{B}\hat{C}&=\frac{1}{4}\text{tr}(\gamma_{\mu}\hat{A}\hat{B}\hat{C})\gamma^{\mu}-\frac{1}{4}\text{tr}(\gamma^5\gamma_{\mu}\hat{A}\hat{B}\hat{C})\gamma^5\gamma^{\mu}\\
&=\hat{A}(BC)-\hat{B}(AC)+\hat{C}(AB)+i\varepsilon_{\mu\nu\lambda\rho}\gamma^5\gamma^{\mu}A^{\nu}B^{\lambda}C^{\rho},
\end{split}
\end{equation*}
where $\gamma^5=i\gamma^0\gamma^1\gamma^2\gamma^3$ and $\varepsilon^{\mu\nu\lambda\rho}$ is the completely antisymmetric tensor with $\varepsilon^{0123}=+1$, which is valid for three arbitrary four-vectors $A^{\mu}$, $B^{\mu}$, and $C^{\mu}$.}. 
The three terms in the third line of Eq. (\ref{Gamma_j}) can be easily reduced to expressions containing at most five gamma matrices:
\begin{equation}
\label{5_gamma_1}
\begin{split}
&L(Ca_j)\hat{G}\hat{\pi}_sR=\left(\hat{a}_j+\frac{G_-}{2Sp'_-\tau'_-}\hat{n}\hat{\Delta}_s\hat{a}_j
-\frac{G_-}{2Sp_-\tau_-}\hat{a}_j\hat{n}\hat{\Delta}_u\right)\hat{G}\hat{\pi}_s\\
&+\left(\hat{a}_j+\frac{G_-}{2Sp'_-\tau'_-}\hat{n}\hat{\Delta}_s\hat{a}_j\right)\left(p'_-\hat{G}-G_-\hat{\pi}_s\right)\frac{\hat{\Delta}_u}{p_-}\\
&+\frac{\hat{a}_j\hat{n}}{2p_-\tau_-}[2(G_-(\Delta_u\pi_s)-p'_-(\Delta_uG))\hat{\Delta}_u-(G_-\Delta_u^2+2\tau_-(\Delta_uG))\hat{\pi}_s+(p'_-\Delta_u^2+2\tau_-(\Delta_u\pi_s))\hat{G}],
\end{split}
\end{equation}
\begin{equation}
\label{5_gamma_2}
\begin{split}
&L\hat{\pi}_u\hat{G}(Ca_j)R=\hat{\pi}_u\hat{G}\left(\hat{a}_j+\frac{G_-}{2Sp'_-\tau'_-}\hat{n}\hat{\Delta}_s\hat{a}_j
-\frac{G_-}{2Sp_-\tau_-}\hat{a}_j\hat{n}\hat{\Delta}_u\right)\\
&+\frac{\hat{\Delta}_s}{p'_-}\left(p_-\hat{G}-G_-\hat{\pi}_u\right)\left(\hat{a}_j-\frac{G_-}{2Sp_-\tau_-}\hat{a}_j\hat{n}\hat{\Delta}_u\right)\\
&+[2(G_-(\Delta_s\pi_u)-p_-(\Delta_sG))\hat{\Delta}_s-(G_-\Delta_s^2+2\tau'_-(\Delta_sG))\hat{\pi}_u+(p_-\Delta_s^2+2\tau'_-(\Delta_s\pi_u))\hat{G}]\frac{\hat{n}\hat{a}_j}{2p'_-\tau'_-},
\end{split}
\end{equation}
\begin{equation}
\label{5_gamma_3}
\begin{split}
&L\hat{G}\left(\hat{a}_j+\frac{\hat{a}_j\hat{\Delta}_s\hat{n}}{2\tau'_-}-\frac{\hat{\Delta}_u\hat{n}\hat{a}_j}{2\tau_-}\right)\hat{G}R=\frac{G_-}{2}\left(\frac{\hat{G}\hat{a}_j\hat{\Delta}_s\hat{n}\hat{\Delta}_u}{p_-\tau'_-}+\frac{\hat{\Delta}_s\hat{n}\hat{\Delta}_u\hat{a}_j\hat{G}}{p'_-\tau_-}\right)+G_-\left(\frac{\hat{a}_j\hat{\Delta}_s\hat{G}}{\tau'_-}+\frac{\hat{G}\hat{\Delta}_u\hat{a}_j}{\tau_-}\right)\\
&+\left[(p'_-+\tau'_-)(Ga_j)+G_-(\Delta_sa_j)\right]\frac{\hat{\Delta}_s\hat{n}\hat{G}}{p'_-\tau'_-}+\left[(p_-+\tau_-)(Ga_j)+G_-(\Delta_ua_j)\right]\frac{\hat{G}\hat{n}\hat{\Delta}_u}{p_-\tau_-}-G^2\hat{a}_j\\
&-\left[(p'_-+\tau'_-)G^2+\frac{\tau'_-G^2_-}{p_-\tau_-}\Delta^2_u\right]\frac{\hat{\Delta}_s\hat{n}\hat{a}_j}{2p'_-\tau'_-}-\left[(p_-+\tau_-)G^2+\frac{\tau_-G^2_-}{p'_-\tau'_-}\Delta^2_s\right]\frac{\hat{a}_j\hat{n}\hat{\Delta}_u}{2p_-\tau_-}+2(Ga_j)\hat{G}\\
&-[G_-\Delta_s^2+2p'_-(G\Delta_s)]\frac{\hat{a}_j\hat{n}\hat{G}}{2p'_-\tau'_-}-[G_-\Delta_u^2+2p_-(G\Delta_u)]\frac{\hat{G}\hat{n}\hat{a}_j}{2p_-\tau_-}\\
&+\frac{G_-}{p_-p'_-}\left[(Ga_j)+\frac{G_-}{\tau'_-}(\Delta_sa_j)+\frac{G_-}{\tau_-}(\Delta_ua_j)\right]\hat{\Delta}_s\hat{n}\hat{\Delta}_u-G^2\left[\frac{(\Delta_sa_j)}{\tau'_-}+\frac{(\Delta_ua_j)}{\tau_-}\right]\hat{n},
\end{split}
\end{equation}

Below, we will further investigate the structure of the vertex correction and discuss its divergences.

\section{Gauge-invariance properties of the vertex-correction function}
\label{VC_GI}
The first aspect we would like to discuss is about the gauge invariance of the expression of $\Gamma_{s,s',l}(p,p',q)$ obtained above. On the one hand, it is clear that $\Gamma_{s,s',l}(p,p',q)$ is invariant under a gauge transformation of the plane wave four-vector potential, as it can be proved by replacing $A^{\mu}(\phi)$ with $A^{\mu}(\phi)+\partial^{\mu}f(\phi)=A^{\mu}(\phi)+n^{\mu}f'(\phi)$, with $f(\phi)$ being an arbitrary function of $\phi$ [we recall, in particular, that $\pi^{\mu}_p(\phi)$ is the kinetic four-momentum of an electron in a plane wave and it is therefore gauge invariant, see Eq. (\ref{pi})]. Now, concerning a gauge transformation of the radiation field and, in particular, of the external photon, we have already discussed that $\Gamma^{\mu}(p,p',q;\phi)$ is written in a form that automatically fulfills the Ward identity, in such a way that one-loop radiative corrections are gauge invariant. In addition, we study here the effect of the additional term $\delta\Gamma^{(\xi)}_{s,s',l}(p,p',q)$ brought about by considering the photon propagator $D^{(\xi)\,\lambda\nu}(x)$ \cite{Itzykson_b_1980}
\begin{equation}
D^{(\xi)\,\lambda\nu}(x)=\int\frac{d^4k}{(2\pi)^4}\frac{e^{-i(kx)}}{k^2-\kappa^2+i0}\left[\eta^{\lambda\nu}+\left(1-\frac{1}{\xi}\right)\frac{k^{\lambda}k^{\nu}}{k^2-\kappa^2+i0}\right],
\end{equation}
in an arbitrary gauge parametrized by the constant $\xi$ (the Lorenz gauge corresponds to $\xi=1$). It is clear from Eq. (\ref{Gamma_2}) that
\begin{equation}
\begin{split}
\delta\Gamma^{(\xi)}_{s,s',l}(p,p',q)=&-ie^2\left(1-\frac{1}{\xi}\right)\int d^4x \int\frac{d^4k}{(2\pi)^4}\,\frac{1}{(k^2-\kappa^2+i0)^2}\\
&\times\bar{U}_{s'}(p',x)\hat{k}\frac{1}{\hat{\Pi}(\phi)+\hat{k}-m+i0}e^{i(qx)}\hat{e}^*_l(q)\frac{1}{\hat{\Pi}(\phi)+\hat{k}-m+i0}\hat{k}U_s(p,x).
\end{split}
\end{equation}
Now, since $[\hat{\Pi}(\phi)-m]U_s(p,x)=[\hat{\Pi}(\phi)-m]U_{s'}(p',x)=0$, we have that $\bar{U}_{s'}(p',x)\hat{k}=\bar{U}_{s'}(p',x)[\hat{\Pi}(\phi)+\hat{k}-m]$ and analogously $\hat{k}U_s(p,x)=[\hat{\Pi}(\phi)+\hat{k}-m]U_s(p,x)$. Thus, the two electron propagators in $\delta\Gamma^{(\xi)}_{s,s',l}(p,p',q)$ simplify and this quantity can be written in the form
\begin{equation}
\delta\Gamma^{(\xi)}_{s,s',l}(p,p',q)=Z^{(\xi)}\int d^4x\,e^{i(qx)}\bar{U}_{s'}(p',x)\hat{e}^*_l(q)U_s(p,x),
\end{equation}
with
\begin{equation}
Z^{(\xi)}=-ie^2\left(1-\frac{1}{\xi}\right)\int\frac{d^4k}{(2\pi)^4}\,\frac{1}{(k^2-\kappa^2+i0)^2}
\end{equation}
being a logarithmically divergent, gauge-dependent constant. However, since $\delta\Gamma^{(\xi)}_{s,s',l}(p,p',q)$ has exactly the same structure of the tree-level matrix element of (virtual) nonlinear Compton scattering, the constant $Z^{(\xi)}$ can be absorbed in the renormalization of the electric charge exactly as in vacuum \cite{Itzykson_b_1980}. Thus, we conclude that the gauge-dependent part of the vertex correction can be absorbed in the renormalization of the electric charge and below we will continue to work in the Lorenz gauge.

\section{Convergence properties of the vertex-correction function}
\label{VC_CP}
Analogously as the corresponding quantity in vacuum, the quantity $\Gamma^{\mu}(p,p',q;\phi)$ is logarithmically divergent in the ultraviolet, as it can be ascertained from the integral in $d^4k$ in Eq. (\ref{Gamma_2}) (see, e.g., the book \cite{Itzykson_b_1980} for the analysis of the vacuum case). Now, if we imagine to expand the exact Volkov propagators in powers of the external field (see also Fig. \ref{FD_VC}), it is clear that, since the divergence of the corresponding vacuum amplitude is logarithmic, all resulting terms depending on the field are ultraviolet convergent because the loop contains at least three vacuum electron propagators apart from the photon propagator. It is important to stress here that this does not imply that the whole field-dependent part of $\Gamma_{s,s',l}(p,p',q)$ is ultraviolet convergent because the terms dependent on the field exclusively through the external electron states are still logarithmically divergent. For this reason, the correct way of regularizing the vertex correction in the plane wave is to regularize the quantity $\Gamma^{\mu}(p,p',q;\phi)$ (see also Ref. \cite{Morozov_1981}). Since the divergence at hand is only logarithmic one first writes $\Gamma^{\mu}(p,p',q;\phi)=\Gamma^{\mu}(p,p',q;\phi)-\Gamma_0^{\mu}(p,p',q)+\Gamma_0^{\mu}(p,p',q)$, where $\Gamma_0^{\mu}(p,p',q)=\Gamma^{\mu}(p,p',q;\phi)|_{A^{\mu}(\phi)=0}$, and notices that $\Gamma^{\mu}(p,p',q;\phi)-\Gamma_0^{\mu}(p,p',q)$ is ultraviolet convergent. Then, one can regularize the vacuum expression $\Gamma_0^{\mu}(p,p',q)$ exactly as in the vacuum, i.e., by subtracting the same expression evaluated for $q^{\mu}=0$ and for $\hat{p}=\hat{p}'=m$ \cite{Itzykson_b_1980} (notice that the conservation laws in a plane wave already imply that $p^{\prime\mu}=p^{\mu}$ because these four-momenta are on-shell). In conclusion, by assuming that $e$ indicates the physical electron charge, we continue by investigating the regularized vertex function $\Gamma_R^{\mu}(p,p',q;\phi)=\Gamma^{\mu}(p,p',q;\phi)-\Gamma_0^{\mu}(p,p,0)|_{\hat{p}=m}$. By using the master integrals
\begin{align}
\int\frac{d^4k}{(2\pi)^4}e^{iSk^2}&=-\frac{i}{16\pi^2S^2},\\
\int\frac{d^4k}{(2\pi)^4}k^{\mu}k^{\nu}e^{iSk^2}&=\frac{\eta^{\mu\nu}}{4}\int\frac{d^4k}{(2\pi)^4}k^2e^{iSk^2}=\frac{\eta^{\mu\nu}}{32\pi^2S^3},
\end{align}
it is straightforward to take the integral in $d^4k$ in $\Gamma_0^{\mu}(p,p,0)|_{\hat{p}=m}$ and to obtain the result
\begin{equation}
\Gamma_0^{\mu}(p,p,0)|_{\hat{p}=m}=-i\frac{\alpha}{2\pi}\gamma^{\mu}\int_0^{\infty}\frac{dsdudt}{S^3}\,e^{-i\kappa^2t-i\frac{(s+u)^2}{S}m^2}\left\{m^2\left[2t-\frac{(s+u)^2}{S}\right]+i\right\}.
\end{equation}
From the derivations, it is clear that $\Gamma_0^{\mu}(p,p,0)|_{\hat{p}=m}$ has only components $\Gamma_{0,-}(p,p,0)|_{\hat{p}=m}$ and $\Gamma_{0,\perp,j}(p,p,0)|_{\hat{p}=m}$, and then that $\Gamma_{R,q}(p,p',q;\phi)=\Gamma_q(p,p',q;\phi)$, which, as we have mentioned, can be shown to vanish for $A^{\mu}(\phi)=0$ (see the appendix).

Now, we would like to investigate the convergence properties of the proper time integrals in $\Gamma_{R,-}(p,p',q;\phi)$ and $\Gamma_{R,\perp,j}(p,p',q;\phi)$. It is first convenient to use the following identity \cite{Schubert_2001}
\begin{equation}
\begin{split}
\int_0^{\infty}ds\int_0^{\infty}du\int_0^{\infty}dt&=\int_0^{\infty}ds\int_0^{\infty}du\int_0^{\infty}dt\int_0^{\infty}dS\,\delta(S-s-u-t)\\
&=\int_0^{\infty}dS\int_0^Sds\int_0^Sdu\int_0^Sdt\,\delta(S-s-u-t)\\
&=\int_0^{\infty}dS\,S^2\int_0^1dx\int_0^1dy\int_0^1dz\,\delta(1-x-y-z),
\end{split}
\end{equation}
where in the last line we performed the changes of variables $s=xS$, $u=yS$, and $t=zS$. By setting
\begin{equation}
\label{int_3d}
\int_{\delta} dxdydz=\int_0^1dx\int_0^1dy\int_0^1dz\,\delta(1-x-y-z),
\end{equation}
it is instructive to report the expression of $\Gamma_0^{\mu}(p,p,0)|_{\hat{p}=m}$ in terms of the new variables:
\begin{equation}
\Gamma_0^{\mu}(p,p,0)|_{\hat{p}=m}=-i\frac{\alpha}{2\pi}\gamma^{\mu}\int_0^{\infty}dS\int_{\delta}dxdydz\,e^{-i\kappa^2zS-im^2(x+y)^2S}\left\{m^2[2z-(x+y)^2]+\frac{i}{S}\right\},
\end{equation}
because it clearly shows that only the term whose integrand is proportional to $i$ is (logarithmically) divergent (in the limit $S\to 0$). This divergence is related with the ultraviolet logarithmic divergence of the vertex-correction function. Keeping in mind that $z=1-x-y$ [see Eq. (\ref{int_3d})], another divergence for $x+y\to 0$ arises for a massless photon ($\kappa^2=0$), which corresponds to the infrared divergence of the vertex-correction function. By means of the above change of variables, we obtain
\begin{equation}
\label{Gamma_R_-}
\begin{split}
&\Gamma_{R,-}(p,p',q;\phi)=\frac{\alpha}{2\pi}\hat{n}\int_0^{\infty}\frac{dS}{S}\int_{\delta} dxdydz\,e^{-i\kappa^2zS}\left[e^{-ig^2S}-e^{-im^2(x+y)^2S}\right]\\
&\quad-\frac{i\alpha}{2\pi}\int_0^{\infty}dS\int_{\delta} dxdydz\,e^{-i\kappa^2zS}\Bigg\{e^{-ig^2S}\Bigg[\left(2(\pi_s\pi_u)+g^2\right)\hat{n}-2g_-(\hat{\pi}_sR+L\hat{\pi}_u)-\hat{g}\hat{\pi}_s\hat{n}-\hat{n}\hat{\pi}_u\hat{g}\\
&\left.\quad+2\tau_-L\hat{g}+2\tau'_-\hat{g}R+2g_-\hat{g}-g_-^2\frac{\hat{\Delta}_s\hat{n}\hat{\Delta}_u}{p_-p'_-}\right]-m^2\hat{n}e^{-im^2(x+y)^2S}[2z-(x+y)^2]\Bigg\},
\end{split}
\end{equation}
and
\begin{equation}
\label{Gamma_R_j}
\begin{split}
&\Gamma_{R,\perp,j}(p,p',q;\phi)=-\frac{\alpha}{2\pi}\hat{a}_j\int_0^{\infty}\frac{dS}{S}\int_{\delta} dxdydz\,e^{-i\kappa^2zS}\left[e^{-ig^2S}-e^{-im^2(x+y)^2S}\right]\\
&\quad+\frac{i\alpha}{2\pi}\int_0^{\infty}dS\int_{\delta} dxdydz\,e^{-i\kappa^2zS}\left\langle e^{-ig^2S}\left\{2(\pi_s\pi_u)\left(\hat{a}_j+\frac{g_-}{2p'_-\tau'_-}\hat{n}\hat{\Delta}_s\hat{a}_j-\frac{g_-}{2p_-\tau_-}\hat{a}_j\hat{n}\hat{\Delta}_u\right)\right.\right.\\
&\quad+\frac{i}{S}\left(\frac{g_-}{2p'_-\tau'_-}\hat{n}\hat{\Delta}_s\hat{a}_j-\frac{g_-}{2p_-\tau_-}\hat{a}_j\hat{n}\hat{\Delta}_u\right)-L(Ca_j)\hat{g}\hat{\pi}_sR-L\hat{\pi}_u\hat{g}(Ca_j)R\\
&\quad-L\hat{g}\left(\hat{a}_j+\frac{\hat{a}_j\hat{\Delta}_s\hat{n}}{2\tau'_-}-\frac{\hat{\Delta}_u\hat{n}\hat{a}_j}{2\tau_-}\right)\hat{g}R-2S(gg_1)\left(\frac{\tau_-}{p'_-}\hat{n}\hat{\Delta}_s\hat{a}_j-\frac{\tau'_-}{p_-}\hat{a}_j\hat{n}\hat{\Delta}_u+2g_-\hat{a}_j-2(ga_j)\hat{n}\right.\\
&\quad+\hat{n}\hat{a}_j\hat{\pi}_s+\hat{\pi}_u\hat{a}_j\hat{n}\bigg)+2i\left((g_1a_j)+x(\mathcal{A}'_sa_j)-y(\mathcal{A}'_ua_j)\right)\hat{n}+i\left[x-(y+z)\frac{\tau_-}{p'_-}\right]\hat{\mathcal{A}}_s'\hat{n}\hat{a}_j\\
&\quad-i\left[y-(x+z)\frac{\tau'_-}{p_-}\right]\hat{a}_j\hat{n}\hat{\mathcal{A}}_u'\Bigg\}-m^2\hat{a}_je^{-im^2(x+y)^2S}[2z-(x+y)^2]\Bigg\rangle,
\end{split}
\end{equation}
where
\begin{equation}
\begin{split}
&L(Ca_j)\hat{g}\hat{\pi}_sR=\left(\hat{a}_j+\frac{g_-}{2p'_-\tau'_-}\hat{n}\hat{\Delta}_s\hat{a}_j
-\frac{g_-}{2p_-\tau_-}\hat{a}_j\hat{n}\hat{\Delta}_u\right)\hat{g}\hat{\pi}_s\\
&+\left(\hat{a}_j+\frac{g_-}{2p'_-\tau'_-}\hat{n}\hat{\Delta}_s\hat{a}_j\right)\left(p'_-\hat{g}-g_-\hat{\pi}_s\right)\frac{\hat{\Delta}_u}{p_-}\\
&+\frac{\hat{a}_j\hat{n}}{2p_-\tau_-}[2(g_-(\Delta_u\pi_s)-p'_-(\Delta_ug))\hat{\Delta}_u-(g_-\Delta_u^2+2\tau_-(\Delta_ug))\hat{\pi}_s+(p'_-\Delta_u^2+2\tau_-(\Delta_u\pi_s))\hat{g}],
\end{split}
\end{equation}
\begin{equation}
\begin{split}
&L\hat{\pi}_u\hat{g}(Ca_j)R=\hat{\pi}_u\hat{g}\left(\hat{a}_j+\frac{g_-}{2p'_-\tau'_-}\hat{n}\hat{\Delta}_s\hat{a}_j-\frac{g_-}{2p_-\tau_-}\hat{a}_j\hat{n}\hat{\Delta}_u\right)
\\
&+\frac{\hat{\Delta}_s}{p'_-}\left(p_-\hat{g}-g_-\hat{\pi}_u\right)\left(\hat{a}_j-\frac{g_-}{2p_-\tau_-}\hat{a}_j\hat{n}\hat{\Delta}_u\right)\\
&+[2(g_-(\Delta_s\pi_u)-p_-(\Delta_sg))\hat{\Delta}_s-(g_-\Delta_s^2+2\tau'_-(\Delta_sg))\hat{\pi}_u+(p_-\Delta_s^2+2\tau'_-(\Delta_s\pi_u))\hat{g}]\frac{\hat{n}\hat{a}_j}{2p'_-\tau'_-},
\end{split}
\end{equation}
\begin{equation}
\begin{split}
&L\hat{g}\left(\hat{a}_j+\frac{\hat{a}_j\hat{\Delta}_s\hat{n}}{2\tau'_-}-\frac{\hat{\Delta}_u\hat{n}\hat{a}_j}{2\tau_-}\right)\hat{g}R=\frac{g_-}{2}\left(\frac{\hat{g}\hat{a}_j\hat{\Delta}_s\hat{n}\hat{\Delta}_u}{p_-\tau'_-}+\frac{\hat{\Delta}_s\hat{n}\hat{\Delta}_u\hat{a}_j\hat{g}}{p'_-\tau_-}\right)+g_-\left(\frac{\hat{a}_j\hat{\Delta}_s\hat{g}}{\tau'_-}+\frac{\hat{g}\hat{\Delta}_u\hat{a}_j}{\tau_-}\right)\\
&+\left[(p'_-+\tau'_-)(ga_j)+g_-(\Delta_sa_j)\right]\frac{\hat{\Delta}_s\hat{n}\hat{g}}{p'_-\tau'_-}+\left[(p_-+\tau_-)(ga_j)+g_-(\Delta_ua_j)\right]\frac{\hat{g}\hat{n}\hat{\Delta}_u}{p_-\tau_-}-g^2\hat{a}_j\\
&-\left[(p'_-+\tau'_-)g^2+\frac{\tau'_-g^2_-}{p_-\tau_-}\Delta^2_u\right]\frac{\hat{\Delta}_s\hat{n}\hat{a}_j}{2p'_-\tau'_-}-\left[(p_-+\tau_-)g^2+\frac{\tau_-g^2_-}{p'_-\tau'_-}\Delta^2_s\right]\frac{\hat{a}_j\hat{n}\hat{\Delta}_u}{2p_-\tau_-}+2(ga_j)\hat{g}\\
&-[g_-\Delta_s^2+2p'_-(g\Delta_s)]\frac{\hat{a}_j\hat{n}\hat{g}}{2p'_-\tau'_-}-[g_-\Delta_u^2+2p_-(g\Delta_u)]\frac{\hat{g}\hat{n}\hat{a}_j}{2p_-\tau_-}\\
&+\frac{g_-}{p_-p'_-}\left[(ga_j)+\frac{g_-}{\tau'_-}(\Delta_sa_j)+\frac{g_-}{\tau_-}(\Delta_ua_j)\right]\hat{\Delta}_s\hat{n}\hat{\Delta}_u-g^2\left[\frac{(\Delta_sa_j)}{\tau'_-}+\frac{(\Delta_ua_j)}{\tau_-}\right]\hat{n},
\end{split}
\end{equation}
and where it is clear that also in the case of $\Gamma_{R,\perp,j}(p,p',q;\phi)$ the only term requiring regularization is the one analogous to that in the first line of Eq. (\ref{Gamma_R_-}). Due to the above change of variables, the various quantities appearing in $\Gamma_{R,-}(p,p',q;\phi)$, and $\Gamma_{R,\perp,j}(p,p',q;\phi)$ have to be interpreted as
\begin{align}
\tau'_-&=zp'_--yq_-=(1-x-y)p'_--yq_-, & \pi_s^{\mu}&=\pi^{\mu}_{p'}(\theta'_S), & \Delta^{\mu}_s &=\mathcal{A}^{\mu}(\theta'_S)-\mathcal{A}^{\mu}(\phi), \\
\tau_-&=zp_-+xq_-=(1-x-y)p_-+xq_-, & \pi_u^{\mu}&=\pi^{\mu}_p(\theta_S), & \Delta^{\mu}_u&=\mathcal{A}^{\mu}(\theta_S)-\mathcal{A}^{\mu}(\phi),
\end{align}
where
\begin{align}
\theta'_S&=\phi+2x\tau'_-S=\phi+2x[(1-x-y)p'_--yq_-]S,\\
\theta_S&=\phi-2y\tau_-S=\phi-2y[(1-x-y)p_-+xq_-]S.
\end{align}

The formal definitions of the other quantities like $L$, $R$, $C^{\mu}$, $Q^{\lambda}$, and $Q^{\prime\,\lambda}$ remain unchanged and the additional quantities
\begin{equation}
g^{\mu}=\frac{G^{\mu}}{S}=x\int_0^1d\eta\,\pi^{\mu}_{p'}(\theta'_{\eta S})+y\int_0^1d\eta\,\pi^{\mu}_p(\theta_{\eta S})
\end{equation}
and
\begin{equation}
\begin{split}
g_1^{\mu}&=\frac{G_1^{\mu}}{S^2}=\frac{d}{d\phi}\left[x^2\int_0^1d\eta\,\eta\pi^{\mu}_{p'}(\theta'_{\eta S})-y^2\int_0^1d\eta\,\eta\pi^{\mu}_p(\theta'_{\eta S})\right]\\
&=\frac{x}{2\tau'_-S}\left[\pi^{\mu}_{p'}(\theta'_S)-\int_0^1d\eta\,\pi^{\mu}_{p'}(\theta'_{\eta S})\right]+\frac{y}{2\tau_-S}\left[\pi^{\mu}_p(\theta_S)-\int_0^1d\eta\,\pi^{\mu}_p(\theta_{\eta S})\right],
\end{split}
\end{equation}
which is regular in the limit $S\to 0$ (and also in the limits $\tau_-\to 0$ and $\tau'_-\to 0$), have been also introduced.

\section{The locally-constant field approximation}
\label{VC_LCFA}
In this section, we would like to investigate the regularized vertex-correction function $\Gamma_R^{\mu}(p,p',q;\phi)$ in the so-called locally-constant field approximation (LCFA) \cite{Reiss_1962,Ritus_1985,Baier_b_1998,Di_Piazza_2012}. Under this approximation quantum processes in an external field are assumed to form over a length much shorter than the typical length where the external field significantly varies \cite{Reiss_1962,Ritus_1985,Baier_b_1998,Di_Piazza_2012}. As a general condition of validity of the LCFA in a plane wave, one assumes that the strength of the vector potential of the plane wave times the elementary charge is much larger than the electron mass. This condition is based on the idea that the strength of the vector potential scales as the strength of the electric field of the wave times the typical field wavelength, and that the LCFA applies for larger and larger wavelengths (see Refs. \cite{Baier_1989,Khokonov_2002,Di_Piazza_2007, Wistisen_2015,Harvey_2015,Dinu_2016,Di_Piazza_2018_c,Alexandrov_2019,Di_Piazza_2019, Ilderton_2019_b,Podszus_2019,Ilderton_2019,Raicher_2020} for more refined results and investigations about the validity of the LCFA). In order to study the structure of the vertex-correction function $\Gamma_R^{\mu}(p,p',q;\phi)$, it is first useful to exploit the general structure of the external plane wave, in particular, to rewrite the phase $G^2/S$ in a convenient form [see Eqs. (\ref{Gamma_-}) and (\ref{Gamma_j})]. By starting from the identity $v^2=2v_+v_--\bm{v}^2_{\perp}$, valid for a generic four-vector $v^{\mu}$, it can easily be shown that
\begin{equation}
\label{G^2}
\begin{split}
G^2&=\frac{s(p'_-s+p_-u)}{p'_-}(m^2+\delta m_s^2)+\frac{u(p'_-s+p_-u)}{p_-}(m^2+\delta m_u^2)\\
&\quad+usp_-p'_-\left\{\frac{1}{p'_-}\left[\bm{\pi}_{p',\perp}(\phi)-\frac{1}{s}\int_0^sds'\bm{\Delta}_{s',\perp}\right]-\frac{1}{p_-}\left[\bm{\pi}_{p,\perp}(\phi)-\frac{1}{u}\int_0^udu'\bm{\Delta}_{u',\perp}\right]\right\}^2,
\end{split}
\end{equation}
where we have introduced the laser-induced square mass corrections
\begin{align}
\label{m_corr_s}
\delta m_s^2&=\frac{1}{s}\int_0^sds'\bm{\mathcal{A}}^2_{\perp}(\psi_{s'})-\frac{1}{s^2}\left[\int_0^sds'\bm{\mathcal{A}}_{\perp}(\psi_{s'})\right]^2=\frac{1}{s}\int_0^sds'\bm{\Delta}^2_{s',\perp}-\frac{1}{s^2}\left(\int_0^sds'\bm{\Delta}_{s',\perp}\right)^2,\\
\label{m_corr_u}
\delta m_u^2&=\frac{1}{u}\int_0^udu'\bm{\mathcal{A}}^2_{\perp}(\psi_{u'})-\frac{1}{u^2}\left[\int_0^udu'\bm{\mathcal{A}}_{\perp}(\psi_{u'})\right]^2=\frac{1}{u}\int_0^udu'\bm{\Delta}^2_{u',\perp}-\frac{1}{u^2}\left(\int_0^udu'\bm{\Delta}_{u',\perp}\right)^2.
\end{align}
We notice that for the present case of on-shell electrons, the three quantity $G^2$ is non-negative, a property which will be used below. Also, we observe that for the evaluation of $\Gamma_R^{\mu}(p,p',q;\phi)$ the phase $\phi$ is fixed but the vector potential depends on the integration variables $s$, $u$, and $t$ (or $x$, $y$, and $S$). Thus, within the integration region, the terms in the phases depending on the vector potential become larger and larger, leading in turn to highly-oscillating integrands. Thus, the largest contributions to the integrals in the proper times come from the regions where these variables are sufficiently small that the squares of the mass corrections $\delta m_s^2$, and $\delta m_u^2$ are of the order of $m^2$ \cite{Di_Piazza_2013} (see also Ref. \cite{Meuren_2015b} for a study of the subleading contributions arising from the saddle points of the phases). In order to implement this idea, we assume that the variables $s$ and $u$ in Eqs. (\ref{m_corr_s})-(\ref{m_corr_u}) in the regions mainly contributing to the corresponding integrals are sufficiently small to expand the integrands in those equations for $\psi_{s'}$, and $\psi_{u'}$ around $\phi$ (the validity of this assumption is checked \textit{a posteriori}). It is appropriate to perform the expansions up to terms proportional to the second derivative of $\bm{\mathcal{A}}_{\perp}(\phi)$ because the leading-order contributions to $\delta m_s^2$, and to $\delta m_u^2$ turn out to be proportional to $\bm{\mathcal{A}}_{\perp}^{\prime\,2}(\phi)$, i.e., to the square of the first-order correction. Indeed, all contributions proportional to $\bm{\mathcal{A}}''_{\perp}(\phi)$ cancel out and one obtains
\begin{align}
\label{m_corr_LCFA}
\delta m_s^2&\approx \frac{1}{3}m^2\left[\frac{t\chi_{p'}(\phi)-u\chi_q(\phi)}{S}\right]^2m^4s^2,&&
\delta m_u^2\approx \frac{1}{3}m^2\left[\frac{t\chi_p(\phi)+s\chi_q(\phi)}{S}\right]^2m^4u^2,
\end{align}
where $\chi_p(\phi)=p_-|\bm{\mathcal{E}}_{\perp}(\phi)|/m^3$, $\chi_{p'}(\phi)=p'_-|\bm{\mathcal{E}_{\perp}}(\phi)|/m^3$, and $\chi_q(\phi)=q_-|\bm{\mathcal{E}}_{\perp}(\phi)|/m^3=\chi_{p'}(\phi)-\chi_{p'}(\phi)$, with $\bm{\mathcal{E}}_{\perp}(\phi)=-\bm{\mathcal{A}}'_{\perp}(\phi)$ (recall that we have assumed that $q_-> 0$, such that $\chi_q(\phi)\ge 0$). Now, in order to obtain the range of validity of the above approximations, it is easier to consider the typical situation in which $p_-\sim p'_-\sim q_-$ and to indicate as $p_{0,-}$ this common light-cone energy scale. Correspondingly, by indicating as $E_0$ and $\omega_0$, the amplitude and the typical angular frequency of the background plane wave ($\omega_0$ can also be thought as the inverse of the typical time interval over which the background field varies significantly), we construct the well-known Lorentz- and gauge-invariant parameters $\xi_0=\mathcal{E}_0/m\omega_0$, $\chi_0=p_{0,-}\mathcal{E}_0/m^3$ and $\eta_0=\chi_0/\xi_0=\omega_0p_{0,-}/m^2$ \cite{Ritus_1985,Baier_b_1998,Di_Piazza_2012}, with $\mathcal{E}_0=|e|E_0$. The above approximations are all valid if the integrals are formed over regions of $s$ and $u$ such that $\omega_0sp_{0,-}=m^2s\eta_0\ll 1$ and $\omega_0up_{0,-}=m^2u\eta_0\ll 1$ (note that the additional proper time variable $t$ appears in the equations in a way that the relevant conditions and estimates do not involve it). Now, from the expressions of the mass corrections within the LFCA, it is easily seen that if $\chi_0\sim 1$ ($\chi_0\gg 1$), then the integrals are formed over the region where $s,u\lesssim 1/m^2$ ($s,u\lesssim 1/\chi_0^{2/3}m^2$). Since here we are interested in situations where $\chi_0\gtrsim 1$, we can for simplicity use the single expression $s,u\lesssim 1/\chi_0^{2/3}m^2$, such that the LCFA is valid if $\eta_0/\chi_0^{2/3}=\chi_0^{1/3}/\xi_0\ll 1$ \cite{Baier_1989,Khokonov_2002,Di_Piazza_2007,Dinu_2016,Di_Piazza_2018_c,Di_Piazza_2019,Ilderton_2019_b,Podszus_2019,Ilderton_2019}. By expanding also the terms in the second line of Eq. (\ref{G^2}) up to the second derivative of $\bm{A}_{\perp}(\phi)$, we obtain that at the leading order in the LCFA the phase $G^2/S$ reads
\begin{equation}
\label{Phase_LCFA_1}
\begin{split}
\frac{G^2}{S}&=g^2S\approx\frac{p'_-s+p_-u}{p'_-S}m^2s\left\{1+\frac{1}{3}\left[\frac{t\chi_{p'}(\phi)-u\chi_q(\phi)}{S}\right]^2m^4s^2\right\}\\
&\quad+\frac{p'_-s+p_-u}{p_-S}m^2u\left\{1+\frac{1}{3}\left[\frac{t\chi_p(\phi)+s\chi_q(\phi)}{S}\right]^2m^4u^2\right\}\\
&\quad+\frac{usp_-p'_-}{S}\left\{\frac{1}{p'_-}\left[\bm{p}'_{\perp}-\bm{\mathcal{A}}_{\perp}(\phi)+m^3s\frac{t\bm{\chi}_{\perp,p'}(\phi)-u\bm{\chi}_{\perp,q}(\phi)}{S}\right]\right.\\
&\left.\qquad-\frac{1}{p_-}\left[\bm{p}_{\perp}-\bm{\mathcal{A}}_{\perp}(\phi)-m^3u\frac{t\bm{\chi}_{\perp,p}(\phi)+s\bm{\chi}_{\perp,q}(\phi)}{S}\right]\right\}^2\\
&\quad+\frac{4}{3}\frac{usp_-p'_-}{S}\left(\frac{s^2\tau_-^{\prime\,2}}{p'_-}-\frac{u^2\tau_-^2}{p_-}\right)\bm{\mathcal{E}}_{\perp}'(\phi)\cdot\bm{\mathcal{V}}_{\perp}(\phi),
\end{split}
\end{equation}
where $\bm{\chi}_{\perp,p}(\phi)=p_-\bm{\mathcal{E}}_{\perp}(\phi)/m^3$ ($\chi_p(\phi)=|\bm{\chi}_{\perp,p}(\phi)|$), $\bm{\chi}_{\perp,p'}(\phi)=p'_-\bm{\mathcal{E}}_{\perp}(\phi)/m^3$ ($\chi_{p'}(\phi)=|\bm{\chi}_{\perp,p'}(\phi)|$), $\bm{\chi}_{\perp,q}(\phi)=q_-\bm{\mathcal{E}}_{\perp}(\phi)/m^3=\bm{\chi}_{\perp,p}(\phi)-\bm{\chi}_{\perp,p'}(\phi)$ ($\chi_q(\phi)=|\bm{\chi}_{\perp,q}(\phi)|$), and where
\begin{equation}
\bm{\mathcal{V}}_{\perp}(\phi)=\frac{1}{p'_-}\left[\bm{p}'_{\perp}-\bm{\mathcal{A}}_{\perp}(\phi)\right]-\frac{1}{p_-}\left[\bm{p}_{\perp}-\bm{\mathcal{A}}_{\perp}(\phi)\right].
\end{equation}
The appearance of $\bm{\mathcal{A}}_{\perp}(\phi)$ in the vector $\bm{\mathcal{V}}_{\perp}(\phi)$ seems to indicate that indeed the last line of Eq. (\ref{Phase_LCFA_1}) is leading order in the LCFA because $|\bm{\mathcal{V}}_{\perp}(\phi)|$ scales as $1/\omega_0$. We should however recall that the final object to be computed is $\Gamma_{s,s',l}(p,p',q)$ in Eq. (\ref{Gamma^mu}). Now, if we compute the phase $\Phi(p,p',q;\phi)$ resulting from the functions in Eq. (\ref{Gamma^mu}) other than $\Gamma^{\mu}(p,p',q;\phi)$, after taking the integrals in $\bm{x}_{\perp}$ and $T$, we obtain (apart from an inessential constant)
\begin{equation}
\label{Phi_NCS}
\Phi(p,p',q;\phi)=\frac{q_-}{2p_-p'_-}\int_0^{\phi}d\phi'\left\{m^2+\frac{p_-p'_-}{q_-^2}q^2+\left[\bm{p}_{\perp}-\bm{\mathcal{A}}_{\perp}(\phi')-\frac{p_-}{q_-}\bm{q}_{\perp}\right]^2\right\}
\end{equation}
together with the conservation laws $\bm{p}_{\perp}=\bm{p}'_{\perp}+\bm{q}_{\perp}$ and $p_-=p'_-+q_-$. It is clear that, apart from the term proportional to $q^2$, this is the phase of nonlinear Compton scattering, as given, e.g., in Ref. \cite{Di_Piazza_2018_c}. By using the conservation laws $\bm{p}_{\perp}=\bm{p}'_{\perp}+\bm{q}_{\perp}$ and $p_-=p'_-+q_-$, it is easy to show that
\begin{equation}
\bm{\mathcal{V}}_{\perp}(\phi)=\frac{q_-}{p_-p'_-}\left[\bm{p}_{\perp}-\bm{\mathcal{A}}_{\perp}(\phi)-\frac{p_-}{q_-}\bm{q}_{\perp}\right].
\end{equation}
Since the LCFA corresponds to evaluate the remaining integral in $\phi$ in Eq. (\ref{Gamma^mu}), it is clear that in the region where most of the photons are emitted (and tacitly assuming that the virtuality $q^2$ is less or of the order of $m^2$), it is $|\bm{\mathcal{V}}_{\perp}(\phi)|\sim m$ and the last line in Eq. (\ref{Phase_LCFA_1}) can be neglected. Finally, by applying the changes of variables discussed above, we obtain the final expression of $g^2S$ within the LCFA in the form
\begin{equation}
\label{Phase_LCFA}
\begin{split}
g^2S&\approx\frac{p'_-x+p_-y}{p'_-}xm^2S\left\{1+\frac{1}{3}[z\chi_{p'}(\phi)-y\chi_q(\phi)]^2x^2m^4S^2\right\}\\
&\quad+\frac{p'_-x+p_-y}{p_-}ym^2S\left\{1+\frac{1}{3}[z\chi_p(\phi)+x\chi_q(\phi)]^2y^2m^4S^2\right\}\\
&\quad+xyp_-p'_-S\left\langle\frac{1}{p'_-}\left\{\bm{p}'_{\perp}-\bm{\mathcal{A}}_{\perp}(\phi)+xm^3S[z\bm{\chi}_{\perp,p'}(\phi)-y\bm{\chi}_{\perp,q}(\phi)]\right\}\right.\\
&\left.\qquad-\frac{1}{p_-}\left\{\bm{p}_{\perp}-\bm{\mathcal{A}}_{\perp}(\phi)-ym^3S[z\bm{\chi}_{\perp,p}(\phi)+x\bm{\chi}_{\perp,q}(\phi)]\right\}\right\rangle^2,
\end{split}
\end{equation}
where $z=1-x-y$.

Finally, we point out that have explicitly proved that in the constant-crossed field case $\bm{\mathcal{A}}_{\perp}(\phi)=-\bm{\mathcal{E}}_0\phi$, the phase $g^2S$ reduces to the phase computed in Ref.~\cite{Morozov_1981}. The same can be verified starting from Eq. (\ref{Phase_LCFA}) by setting $\bm{\mathcal{E}}_{\perp}(\phi)=\bm{\mathcal{E}}_0$ and we note that our expression of the phase of the vertex-correction function is not only more general but also much more compact than that presented in Ref.~\cite{Morozov_1981}. A comparison of the final expression of the pre-exponent was not carried out as the form presented in Ref.~\cite{Morozov_1981} has a very different structure from ours, due to employed transformations there, which are appropriate only to the constant-crossed field case.

Passing now to the pre-exponents of the components $\Gamma_{R,-}(p,p',q;\phi)$ [see Eq. (\ref{Gamma_R_-})] and $\Gamma_{R,\perp,j}(p,p',q;\phi)$ [see Eq. (\ref{Gamma_R_j})], one has to expand the field-dependent terms in Eqs. (\ref{Gamma_R_-})-(\ref{Gamma_R_j}) around the phase $\phi$. Taking into account that the final quantity to be evaluated is $-ie\Gamma_{s,s',l}(p,p',q)$ in Eq. (\ref{Gamma^mu}), a lengthy by straightforward calculation shows that
\begin{equation}
\label{Gamma_R_-_LCFA}
\begin{split}
&\Gamma_{R,-}(p,p',q;\phi)\approx\frac{\alpha}{2\pi}\hat{n}\int_0^{\infty}\frac{dS}{S}\int_{\delta} dxdydz\,e^{-i\kappa^2zS}\left[e^{-ig^2S}-e^{-im^2(x+y)^2S}\right]\\
&\quad-\frac{i\alpha}{2\pi}\int_0^{\infty}dS\int_{\delta} dxdydz\,e^{-i\kappa^2zS}\Bigg\langle e^{-ig^2S}\Bigg\{(p_-+p'_--g_-)\left(\frac{1-x}{p'_-}+\frac{1-y}{p_-}\right)m^2\hat{n}-2m^2\hat{n}\\
&\quad+2m[(p_-+p'_--g_-)(x+y)-g_-]+[(2-x)(p_-b'+p'_-b)-g_-b][(2-y)(p_-b'+p'_-b)-g_-b']\frac{\bm{\mathcal{A}}^{\prime\,2}_{\perp}}{p_-p'_-}\hat{n}\\
&\quad+(1-x)(1-y)p_-p'_-\bm{\mathcal{V}}_{\perp}^2\hat{n}-2(1-x)(1-y)p_-p'_-\left(\frac{2-x}{1-x}\frac{b'}{p'_-}+\frac{2-y}{1-y}\frac{b}{p_-}\right)\bm{\mathcal{V}}_{\perp}\cdot\bm{\mathcal{A}}'_{\perp}\hat{n}\\
&\quad+2m\left[g_--(p_-+p'_--g_-)(x+y)\right]\left(\frac{b'}{p'_-}+\frac{b}{p_-}\right)\hat{n}\hat{\mathcal{A}}'\\
&\left.\quad-(p_-+p'_--g_-)\left[x\frac{b'}{p'_-}\hat{\mathcal{A}}'\hat{n}\left(\frac{q_-}{p_-}m+p'_-\bm{\gamma}_{\perp}\cdot\bm{\mathcal{V}}_{\perp}\right)+y\frac{b}{p_-}\left(\frac{q_-}{p'_-}m+p_-\bm{\gamma}_{\perp}\cdot\bm{\mathcal{V}}_{\perp}\right)\hat{n}\hat{\mathcal{A}}'\right]\right\}\\
&\quad-m^2\hat{n}e^{-im^2(x+y)^2S}[2z-(x+y)^2]\Bigg\rangle,
\end{split}
\end{equation}
where $g^2$ is obtained from Eq. (\ref{Phase_LCFA}), $b=y\tau_-S$, $b'=x\tau'_-S$, and where all fields and derivatives are evaluated at $\phi$.

Analogously, one obtains the following expression for $\Gamma_{R,\perp,j}(p,p',q;\phi)$ within the LCFA:
\begin{equation}
\label{Gamma_R_j_LCFA}
\begin{split}
&\Gamma_{R,\perp,j}(p,p',q;\phi)\approx-\frac{\alpha}{2\pi}\hat{a}_j\int_0^{\infty}\frac{dS}{S}\int_{\delta} dxdydz\,e^{-i\kappa^2zS}\left[e^{-ig^2S}-e^{-im^2(x+y)^2S}\right]\\
&\quad+\frac{i\alpha}{2\pi}\int_0^{\infty}dS\int_{\delta} dxdydz\,e^{-i\kappa^2zS}\left\langle e^{-ig^2S}\left\{2[(\pi_s\pi_u)-((\pi_s+\pi_u)g)]\left(\hat{a}_j+\frac{g_-b'}{p'_-\tau'_-}\hat{n}\hat{\mathcal{A}}'\hat{a}_j+\frac{g_-b}{p_-\tau_-}\hat{a}_j\hat{n}\hat{\mathcal{A}}'\right)\right.\right.\\
&\quad+\frac{i}{S}\left(\frac{g_-b'}{p'_-\tau'_-}\hat{n}\hat{\mathcal{A}}'\hat{a}_j+\frac{g_-b}{p_-\tau_-}\hat{a}_j\hat{n}\hat{\mathcal{A}}'\right)+L(Ca_j)\hat{\pi}_s\hat{g}R+L\hat{g}\hat{\pi}_u(Ca_j)R\\
&\quad-L\hat{g}\left(\hat{a}_j+\frac{b'}{\tau'_-}\hat{a}_j\hat{\mathcal{A}}'\hat{n}+\frac{b}{\tau_-}\hat{\mathcal{A}}'\hat{n}\hat{a}_j\right)\hat{g}R-2S(gg_1)\left\{\left[\frac{2b'\tau_-}{p'_-}+(2-y)b+xb'\right]\hat{n}\hat{\mathcal{A}}'\hat{a}_j\right.\\
&\quad+\left[\frac{2b\tau'_-}{p_-}+(2-x)b'+yb\right]\hat{a}_j\hat{n}\hat{\mathcal{A}}'+m\left[\frac{p'_-}{p_-}-\frac{p_-}{p'_-}+x\left(1-\frac{p'_-}{p_-}\right)-y\left(1-\frac{p_-}{p'_-}\right)\right]\hat{n}\hat{a}_j\\
&\quad-p_-(1-y)\bm{\gamma}_{\perp}\cdot\bm{\mathcal{V}}_{\perp}\hat{n}\hat{a}_j-p'_-(1-x)\hat{n}\hat{a}_j\bm{\gamma}_{\perp}\cdot\bm{\mathcal{V}}_{\perp}\bigg\}+i(x-y)(x+y-2)\bm{\mathcal{A}}'_{\perp}\cdot\bm{a}_j\hat{n}\\
&\quad+i\left[x-(y+z)\frac{\tau_-}{p'_-}\right]\hat{\mathcal{A}}'\hat{n}\hat{a}_j-i\left[y-(x+z)\frac{\tau'_-}{p_-}\right]\hat{a}_j\hat{n}\hat{\mathcal{A}}'\Bigg\}-m^2\hat{a}_je^{-im^2(x+y)^2S}[2z-(x+y)^2]\Bigg\rangle,
\end{split}
\end{equation}
where
\begin{equation}
\begin{split}
&(\pi_s\pi_u)-((\pi_s+\pi_u)g)\approx-\frac{p_-+p'_-}{2g_-}g^2-\frac{g_-}{p_-+p'_-}m^2\\
&\quad+\frac{1}{2}\left(1-\frac{g_-}{p_-+p'_-}\right)\left\{\left(\frac{p_-}{p'_-}
+\frac{p'_-}{p_-}\right)m^2+\left[\bm{\mathcal{V}}_{\perp}-2\left(\frac{b'}{p'_-}+\frac{b}{p_-}\right)\bm{\mathcal{A}}'_{\perp}\right]^2p_-p'_-\right\}\\
&\quad-\frac{p_-^2p_-^{\prime\,2}}{2g_-(p_-+p'_-)}\left[(x-y)\bm{\mathcal{V}}_{\perp}-\frac{(xb'+yb)q_-+2xbp'_--2yb'p_-}{p_-p'_-}\bm{\mathcal{A}}'_{\perp}\right]^2,
\end{split}
\end{equation}
\begin{equation}
\begin{split}
&L(Ca_j)\hat{\pi}_s\hat{g}R+L\hat{g}\hat{\pi}_u(Ca_j)R-L\hat{g}\left(\hat{a}_j+\frac{b'}{\tau'_-}\hat{a}_j\hat{\mathcal{A}}'\hat{n}+\frac{b}{\tau_-}\hat{\mathcal{A}}'\hat{n}\hat{a}_j\right)\hat{g}R\\
&\quad\approx g^2\left(\hat{a}_j+\frac{g_-b'}{p'_-\tau'_-}\hat{n}\hat{\mathcal{A}}'\hat{a}_j+\frac{g_-b}{p_-\tau_-}\hat{a}_j\hat{n}\hat{\mathcal{A}}'\right)+2\bigg[\left(\left(\pi_s+\pi_u-g\right)a_j\right)\\
&\quad-S(p_-+p'_--g_-)(x-y)\bm{\mathcal{A}}'_{\perp}\cdot\bm{a}_j\bigg]L\hat{g}R-2m(ga_j)\left[1-\left(\frac{b'}{p'_-}+\frac{b}{p_-}\right)S\hat{n}\hat{\mathcal{A}}'\right]\\
&\quad-mS(x\hat{a}_j\hat{\mathcal{A}}'-y\hat{\mathcal{A}}'\hat{a}_j)\hat{n}\left[\left(by-b'x-2\frac{bg_-}{p_-}\right)\hat{\mathcal{A}}'+m\left(y+x\frac{p'_-}{p_-}\right)-xp'_-\bm{\gamma}_{\perp}\cdot\bm{\mathcal{V}}_{\perp}\right]\\
&\quad-mS\left[\left(by-b'x+2\frac{b'g_-}{p'_-}\right)\hat{\mathcal{A}}'+m\left(x+y\frac{p_-}{p'_-}\right)+yp_-\bm{\gamma}_{\perp}\cdot\bm{\mathcal{V}}_{\perp}\right]\hat{n}(x\hat{a}_j\hat{\mathcal{A}}'-y\hat{\mathcal{A}}'\hat{a}_j)\\
&\quad-2i(p_-+p'_--g_-)(x+y)S\varepsilon^{\lambda\mu\nu\rho}n_{\lambda}\mathcal{A}'_{\mu}a_{j,\nu}\tilde{n}_{\rho}\gamma^5L\hat{g}R,
\end{split}
\end{equation}
\begin{equation}
\begin{split}
L\hat{g}R&\approx m(x+y)\left[1-\left(\frac{b}{p_-}
+\frac{b'}{p'_-}\right)\hat{n}\hat{\mathcal{A}}'\right]-\frac{xb'}{2p'_-}\hat{\mathcal{A}}'\hat{n}\left(m\frac{q_-}{p_-}+p'_-\bm{\mathcal{V}}_{\perp}\cdot\bm{\gamma}_{\perp}\right)\\
&\quad-\frac{yb}{2p_-}\left(m\frac{q_-}{p'_-}+p_-\bm{\mathcal{V}}_{\perp}\cdot\bm{\gamma}_{\perp}\right)\hat{n}\hat{\mathcal{A}}'-\left(\frac{1}{3}\frac{xb^{\prime\,2}}{p'_-}+\frac{1}{3}\frac{yb^2}{p_-}+\frac{g_-bb'}{p_-p'_-}\right)\hat{n}\bm{\mathcal{A}}_{\perp}^{\prime\,2},
\end{split}
\end{equation}
\begin{equation}
\begin{split}
\gamma^5L\hat{g}R&\approx \gamma^5\left[m(y-x)+m\left(\frac{xb}{p_-}
-\frac{yb'}{p'_-}\right)\hat{n}\hat{\mathcal{A}}'-\frac{xb'}{2p'_-}\hat{\mathcal{A}}'\hat{n}\left(m\frac{q_-}{p_-}+p'_-\bm{\mathcal{V}}_{\perp}\cdot\bm{\gamma}_{\perp}\right)\right.\\
&\quad\left.+\frac{yb}{2p_-}\left(m\frac{q_-}{p'_-}-p_-\bm{\mathcal{V}}_{\perp}\cdot\bm{\gamma}_{\perp}\right)\hat{n}\hat{\mathcal{A}}'-\left(\frac{1}{3}\frac{xb^{\prime\,2}}{p'_-}+\frac{1}{3}\frac{yb^2}{p_-}+\frac{g_-bb'}{p_-p'_-}\right)\hat{n}\bm{\mathcal{A}}_{\perp}^{\prime\,2}\right],
\end{split}
\end{equation}
\begin{equation}
\label{pi_pi_a}
((\pi_s+\pi_u)a_j)\approx-\left[2(b-b')\bm{\mathcal{A}}'_{\perp}+\frac{p_-+p'_-}{q_-}\bm{q}_{\perp}+2\frac{p_-p'_-}{q_-}\bm{\mathcal{V}}_{\perp}\right]\cdot\bm{a}_j,
\end{equation}
\begin{equation}
\label{g_a}
(ga_j)\approx\left[(xb'-yb)\bm{\mathcal{A}}'_{\perp}-(yp_-+xp'_-)\frac{\bm{q}_{\perp}}{q_-}-(x+y)\frac{p_-p'_-}{q_-}\bm{\mathcal{V}}_{\perp}\right]\cdot\bm{a}_j,
\end{equation}
\begin{equation}
\begin{split}
(gg_1)&\approx\left[\frac{x^3b'}{6}+\frac{y^3b}{6}+xy^2p'_-\left(\frac{2b}{3p_-}+\frac{b'}{2p'_-}\right)+yx^2p_-\left(\frac{2b'}{3p'_-}+\frac{b}{2p_-}\right)\right]\bm{\mathcal{A}}_{\perp}^{\prime\,2}\\
&\quad-\frac{xy}{2}(yp'_-+xp_-)\bm{\mathcal{V}}_{\perp}\cdot\bm{\mathcal{A}}'_{\perp},
\end{split}
\end{equation}
where $\gamma^5=i\gamma^0\gamma^1\gamma^2\gamma^3$ and $\varepsilon^{\mu\nu\lambda\rho}$ is the completely antisymmetric tensor with $\varepsilon^{0123}=+1$. In the above equations, the quantity $g^2$ is given in Eq. (\ref{Phase_LCFA}) and we have taken into account that finally we need the matrix elements of these matrices between $\bar{U}_{s'}(p',x)$ and $U_s(p,x)$. The appearance of $\bm{q}_{\perp}$ in Eqs. (\ref{pi_pi_a})-(\ref{g_a}) may suggest that large terms (i.e., of the order of $1/\omega_0$) could appear in total probabilities, if one imagines to carry out the integral over the transverse photon momentum, as one has to shift the variable $\bm{q}_{\perp}$ by a vector containing $\bm{\mathcal{A}}_{\perp}(\phi)$ [see Eq. (\ref{Phi_NCS})] in order to perform the resulting Gaussian integral. However, this does not represent a problem because the vector $\bm{q}_{\perp}$ is always multiplied by $\bm{a}_j$ [see Eqs. (\ref{pi_pi_a})-(\ref{g_a})]. Indeed, the first equality in Eq. (\ref{Gamma_exp}) shows that the components $\Gamma_{R,\perp,j}(p,p',q;\phi)$ are only auxiliary quantities, as one finally needs to compute the components $(\Gamma_R(p,p',q;\phi)\Lambda_j)$. Since by definition the vector $\bm{\Lambda}_{\perp,j}$ is perpendicular to $\bm{q}_{\perp}$ [see Eq. (\ref{Lambda_i})], by replacing $a^{\mu}_j$ with $\Lambda_j^{\mu}$ in $\Gamma_{R,\perp,j}(p,p',q;\phi)$, it is easy to see that the $(\Gamma_R(p,p',q;\phi)\Lambda_j)$ depends on the vector $\bm{q}_{\perp}$ only through $\bm{\Lambda}_{\perp,j}$, a quantity which then, due to completeness, drops out once one computes total probabilities.

Finally, we comment on the scaling of the radiative corrections due to the vertex correction at $\chi_0\gg 1$. This is more easily done in the case of $\Gamma_{R,-}(p,p',q;\phi)$ because one knows that the corresponding amplitude in nonlinear Compton scattering is simply proportional to $\bar{u}_{s'}(p')\hat{n}u_s(p)$ (before one regularizes the amplitude by integrating by parts, see, e.g., \cite{Mackenroth_2011}). Now, the structure of the phase in Eq. (\ref{Phase_LCFA}) shows that at large values of $\chi_0$ the main contribution to the integral in $S$ comes from the region $S\lesssim 1/\chi^{2/3}_0$. Thus, the terms in the preexponent in Eq. (\ref{Gamma_R_-_LCFA}) proportional to $p_-^2\bm{A}_{\perp}^{\prime\,2}(\phi)S^2=e^2p_-^2\bm{\mathcal{E}}_{\perp}^2(\phi)S^2$ give rise to the scaling $\alpha\chi_0^{2/3}$ of the radiative corrections in agreement with the results in Ref. \cite{Morozov_1981}.

\section{Conclusions}
\label{VC_Conclusions}

We have computed the general expression of the one-loop vertex correction in an arbitrary plane-wave background field for the case of two on-shell external electrons and an off-shell external photon. By employing the operator technique within the Furry picture, we have obtained a relatively compact expression, which takes into account exactly the background plane-wave field. By showing explicitly that the vertex correction fulfills a generalized Ward identity, we have singled out the corresponding terms invariant under a gauge transformation of the external photon. As expected, the vertex-correction function features an infrared divergence, which is cured by assigning a small, finite mass to the photon. The ultraviolet divergence of the vertex correction has, instead, been shown to be renormalized as in vacuum. 

The important special case of the locally-constant field approximation has been investigated in detail. We have shown that in the high-field regime $\chi_0\gg 1$ the vertex-correction function induces radiative corrections which scale according to the Ritus-Narozhny conjecture as $\alpha\chi^{2/3}_0$, where $\chi_0$ is the amplitude of the quantum nonlinearity parameter.

\acknowledgments{The author would like to thank J. P. Edwards and C. Schubert for insightful discussions. MALL gratefully acknowledges the hospitality of the Max Planck Institute for Nuclear Physics in the initial phase of the project. MALL is grateful to CONACYT for the financial support during this project.}

\appendix
\section{The component $\Gamma_q(p,p',q;\phi)$ of the vertex-correction function}

In this appendix, we evaluate more explicitly the component $\Gamma_q(p,p',q;\phi)$ of the vertex-correction function although we have seen that it does not contribute to any transition matrix element.

\subsection{General structure of $\Gamma_q(p,p',q;\phi)$}
As we have mentioned in the main text, from the second equality in Eq. (\ref{Gamma_2}) and from the definition of $\Gamma^{\mu}(p,p',q;\phi)$ in Eq. (\ref{Gamma^mu}), we obtain
\begin{equation}
\Gamma_q(p,p',q;\phi)=-ie^2\int\frac{d^4k}{(2\pi)^4}\,\frac{1}{k^2-\kappa^2+i0}\gamma^{\lambda}\frac{1}{\hat{\Pi}(\phi)+\hat{k}-\hat{q}-m+i0}\hat{q}\frac{1}{\hat{\Pi}(\phi)+\hat{k}-m+i0}\gamma_{\lambda}.
\end{equation}
Now, by writing $\hat{q}=\hat{\Pi}(\phi)+\hat{k}-m-[\hat{\Pi}(\phi)+\hat{k}-\hat{q}-m]$ it is clear that we can express $\Gamma_q(p,p',q;\phi)$ as the difference of two terms containing only one propagator in the plane wave, which significantly simplifies its expression:
\begin{equation}
\Gamma_q(p,p',q;\phi)=-ie^2\int\frac{d^4k}{(2\pi)^4}\,\frac{1}{k^2-\kappa^2+i0}\gamma^{\lambda}\left[\frac{1}{\hat{\Pi}(\phi)+\hat{k}-\hat{q}-m+i0}-\frac{1}{\hat{\Pi}(\phi)+\hat{k}-m+i0}\right]\gamma_{\lambda}.
\end{equation}
At this point, by following exactly the same steps as in the main text [see Eq. (\ref{Gamma_f})], it is easy to obtain the resulting expression
\begin{equation}
\begin{split}
\Gamma_q(p,p',q;\phi)&=\frac{\alpha}{4\pi}\int_0^{\infty}\frac{dsdt}{(s+t)^2}e^{-i\kappa^2t-i\frac{\tilde{G}^2_s}{s+t}}\left\{ 2\left[1-\frac{e\hat{n}[\hat{A}(\tilde{\psi}_{0,s})-\hat{A}(\phi)]}{2p'_-}\right]\left[\hat{\pi}_{p'}(\tilde{\psi}_{0,s})+\frac{\hat{\tilde{G}}_s}{s+t}\right]\right.\\
&\left.\left.\qquad\qquad-\frac{\hat{n}}{(s+t)^2}\frac{d\tilde{G}^2_s}{dk_-}+\frac{e\hat{n}[\hat{A}(\tilde{\psi}_{0,s})-\hat{A}(\phi)]}{\tau'_{0,-}+k_-}\left[\hat{\pi}_{p'}(\tilde{\psi}_{0,s})-\frac{\hat{\tilde{G}}_s}{s+t}\right]\right\}\right\vert_{k_-=0}\\
&\quad-\frac{\alpha}{4\pi}\int_0^{\infty}\frac{dudt}{(u+t)^2}e^{-i\kappa^2t-i\frac{\tilde{G}^2_u}{u+t}}\left\{ 2\left[\hat{\pi}_p(\tilde{\psi}_{0,u})+\frac{\hat{\tilde{G}}_u}{u+t}\right]\left[1+\frac{e\hat{n}[\hat{A}(\tilde{\psi}_{0,u})-\hat{A}(\phi)]}{2p_-}\right]\right.\\
&\left.\left.\qquad\qquad-\frac{\hat{n}}{(u+t)^2}\frac{d\tilde{G}^2_u}{dk_-}-\left[\hat{\pi}_p(\tilde{\psi}_{0,u})-\frac{\hat{\tilde{G}}_u}{u+t}\right]\frac{e\hat{n}[\hat{A}(\tilde{\psi}_{0,u})-\hat{A}(\phi)]}{\tau_{0,-}+k_-}\right\}\right\vert_{k_-=0},
\end{split}
\end{equation}
where
\begin{align}
\tilde{G}^{\mu}_s&=\int_0^sds'\pi^{\mu}_{p'}(\tilde{\psi}_{0,s'}), && \tilde{\psi}_{0,s}=\phi+2s\tau'_{0,-}+2sk_-,&&\tau'_{0,-}=\frac{t}{s+t}p'_-,\\
\tilde{G}^{\mu}_u&=\int_0^udu'\pi^{\mu}_p(\tilde{\psi}_{0,u'}), && \tilde{\psi}_{0,u}=\phi-2u\tau_{0,-}-2uk_-,&&\tau_{0,-}=\frac{t}{u+t}p_-.
\end{align}
Finally, by evaluating the remaining derivatives with respect to $k_-$ as
\begin{align}
\frac{d\tilde{G}^2_s}{dk_-}&=4(\tilde{G}_s\tilde{G}_{1,s}),\\
\frac{d\tilde{G}^2_u}{dk_-}&=4(\tilde{G}_u\tilde{G}_{1,u}),
\end{align}
where
\begin{align}
\tilde{G}^{\mu}_{1,s}&=\int_0^sds'\,s'\pi^{\prime\,\mu}_{p'}(\tilde{\psi}_{0,s'})=\frac{1}{2(\tau'_{0,-}+k_-)}\left[s\pi^{\mu}_{p'}(\tilde{\psi}_{0,s})-\int_0^sds'\pi^{\mu}_{p'}(\tilde{\psi}_{0,s'})\right],\\
\tilde{G}^{\mu}_{1,u}&=-\int_0^udu'\,u'\pi^{\prime\,\mu}_p(\tilde{\psi}_{0,u'})=\frac{1}{2(\tau_{0,-}+k_-)}\left[u\pi^{\mu}_p(\tilde{\psi}_{0,u})-\int_0^udu'\pi^{\mu}_p(\tilde{\psi}_{0,u'})\right],
\end{align}
we obtain
\begin{equation}
\label{Gamma_q}
\begin{split}
\Gamma_q(p,p',q;\phi)&=\frac{\alpha}{2\pi}\int_0^{\infty}\frac{dsdt}{(s+t)^2}e^{-i\kappa^2t-i\frac{G^2_s}{s+t}}\left\{\left(1-\frac{\hat{n}\hat{\Delta}_{0,s}}{2p'_-}\right)\left[\hat{\pi}_{p'}(\psi_{0,s})+\frac{\hat{G}_s}{s+t}\right]\right.\\
&\left.\qquad\qquad-\frac{2\hat{n}}{(s+t)^2}(G_sG_{1,s})+\frac{\hat{n}\hat{\Delta}_{0,s}}{2\tau'_{0,-}}\left[\hat{\pi}_{p'}(\psi_{0,s})-\frac{\hat{G}_s}{s+t}\right]\right\}\\
&\quad-\frac{\alpha}{2\pi}\int_0^{\infty}\frac{dudt}{(u+t)^2}e^{-i\kappa^2t-i\frac{G^2_u}{u+t}}\left\{\left[\hat{\pi}_p(\psi_{0,u})+\frac{\hat{G}_u}{u+t}\right]\left(1+\frac{\hat{n}\hat{\Delta}_{0,u}}{2p_-}\right)\right.\\
&\left.\qquad\qquad-\frac{2\hat{n}}{(u+t)^2}(G_uG_{1,u})-\left[\hat{\pi}_p(\psi_{0,u})-\frac{\hat{G}_u}{u+t}\right]\frac{\hat{n}\hat{\Delta}_{0,u}}{2\tau_{0,-}}\right\},
\end{split}
\end{equation}
where
\begin{equation}
\Delta^{\mu}_{0,s/u}=e[A^{\mu}(\psi_{0,s/u})-A^{\mu}(\phi)],
\end{equation}
and where all the quantities without the tilde are the same as those with the tilde but with $k_-=0$:
\begin{align}
G^{\mu}_s&=\int_0^sds'\pi^{\mu}_{p'}(\psi_{0,s'}), && G^{\mu}_{1,s}=\frac{1}{2\tau'_{0,-}}\left[s\pi_{p'}^{\mu}(\psi_{0,s})-\int_0^sds'\pi_{p'}^{\mu}(\psi_{0,s'})\right], && \psi_{0,s}=\phi+2s\tau'_{0,-},\\
G^{\mu}_u&=\int_0^udu'\pi^{\mu}_p(\psi_{0,u'}), && G^{\mu}_{1,u}=\frac{1}{2\tau_{0,-}}\left[u\pi_p^{\mu}(\psi_{0,u})-\int_0^udu'\pi_p^{\mu}(\psi_{0,u'})\right], && \psi_{0,u}=\phi-2u\tau_{0,-}.
\end{align}

\subsection{Regularization of $\Gamma_q(p,p',q;\phi)$}
Analogously to the other components of the vertex-correction function, the component $\Gamma_q(p,p',q;\phi)$ has in principle to be regularized as it is apparently logarithmically divergent in the ultraviolet. However, in the case $\Gamma_{R,q}(p,p',q;\phi)$, actually, it is not necessary to perform any subtraction of vacuum terms because, as we will show now, it vanishes for $A^{\mu}(\phi)=0$. It is convenient to perform the change of variable $s=x\sigma$ and $t=(1-x)\sigma$ ($u=x\sigma$ and $t=(1-x)\sigma$) in the first (second) integral in Eq. (\ref{Gamma_q}) \footnote{Note that in this section and afterwards the symbol $x$ should not be confused with a spacetime point as it indicates a single, real variable.} and we obtain
\begin{equation}
\label{Gamma_R_q}
\begin{split}
&\Gamma_{R,q}(p,p',q;\phi)=\frac{\alpha}{2\pi}\int_0^{\infty}\frac{d\sigma}{\sigma}\int_0^1dx\,e^{-i\kappa^2(1-x)\sigma}\\
&\quad\times\left\langle e^{-ix^2\sigma g_s^2}\left\{\left(1-\frac{\hat{n}\hat{\Delta}_{0,s}}{2p'_-}\right)[\hat{\pi}_{p'}(\theta'_{0,\sigma})+x\hat{g}_s]-2x^3\sigma(g_sg_{1,s})\hat{n}+\frac{\hat{n}\hat{\Delta}_{0,s}}{2(1-x)p'_-}[\hat{\pi}_{p'}(\theta'_{0,\sigma})-x\hat{g}_s]\right\}\right.\\
&\quad-\left.e^{-ix^2\sigma g_u^2}\left\{[\hat{\pi}_p(\theta_{0,\sigma})+x\hat{g}_u]\left(1+\frac{\hat{n}\hat{\Delta}_{0,u}}{2p_-}\right)-2x^3\sigma(g_ug_{1,u})\hat{n}-[\hat{\pi}_p(\theta_{0,\sigma})-x\hat{g}_u]\frac{\hat{n}\hat{\Delta}_{0,u}}{2(1-x)p_-}\right\}\right\rangle,
\end{split}
\end{equation}
where
\begin{align}
\theta'_{0,\sigma}&=\phi+2x(1-x)p'_-\sigma,\\
\theta_{0,\sigma}&=\phi-2x(1-x)p_-\sigma,
\end{align}
and
\begin{align}
g^{\mu}_s=\frac{G^{\mu}_s}{s}=\int_0^1d\eta\,\pi_{p'}^{\mu}(\theta'_{0,\eta\sigma}),&& g^{\mu}_{1,s}=\frac{G^{\mu}_{1,s}}{s^2}=\frac{1}{2x(1-x)\sigma p'_-}\left[\pi_{p'}^{\mu}(\theta'_{0,\sigma})-\int_0^1d\eta\,\pi_{p'}^{\mu}(\theta'_{0,\eta\sigma})\right],\\ g^{\mu}_u=\frac{G^{\mu}_u}{u}=\int_0^1d\eta\,\pi_p^{\mu}(\theta_{0,\eta \sigma}),&& g^{\mu}_{1,u}=\frac{G^{\mu}_{1,u}}{u^2}=\frac{1}{2x(1-x)\sigma p_-}\left[\pi_p^{\mu}(\theta_{0,\sigma})-\int_0^1d\eta\,\pi_p^{\mu}(\theta_{0,\eta\sigma})\right].
\end{align}
Note that $g_{1,s}^{\mu}$ and $g_{1,u}^{\mu}$ tend to constant values for $\sigma\to 0$. Since $\Delta^{\mu}_{0,s}$ and $\Delta^{\mu}_{0,u}$ vanish at $\sigma=0$, the only problematic terms are those in the square brackets in Eq. (\ref{Gamma_R_q}) containing exclusively $\hat{\pi}_{p'}(\theta'_{0,\sigma})+x\hat{g}_s$ and $\hat{\pi}_p(\theta_{0,\sigma})+x\hat{g}_u$. However, since $\Gamma_q(p,p',q;\phi)$ will finally be multiplied by Volkov states both from the left and on the right, we can replace $\hat{\pi}_{p'}(\theta'_{0,\sigma})+x\hat{g}_s$ and $\hat{\pi}_p(\theta_{0,\sigma})+x\hat{g}_u$ in the terms which do not contain other gamma matrices as
\begin{align}
\hat{\pi}_{p'}(\theta'_{0,\sigma})+x\hat{g}_s&\rightarrow m(1+x)+\hat{\pi}_{p'}(\theta'_{0,\sigma})-\hat{\pi}_{p'}(\phi)+x\int_0^1d\eta\,[\hat{\pi}_{p'}(\theta'_{0,\eta\sigma})-\hat{\pi}_{p'}(\phi)],\\
\hat{\pi}_p(\theta_{0,\sigma})+x\hat{g}_u&\rightarrow m(1+x)+\hat{\pi}_p(\theta_{0,\sigma})-\hat{\pi}_p(\phi)+x\int_0^1d\eta\,[\hat{\pi}_p(\theta_{0,\eta\sigma})-\hat{\pi}_p(\phi)].
\end{align}
In this way, the only remaining divergent terms are those proportional to $m(1+x)$ but these divergences cancel each other in Eq. (\ref{Gamma_R_q}), which can be conveniently written in the manifestly convergent form
\begin{equation}
\begin{split}
\Gamma_{R,q}(p,p',q;\phi)&=\frac{\alpha}{2\pi}\int_0^{\infty}\frac{d\sigma}{\sigma}\int_0^1dx\,e^{-i\kappa^2(1-x)\sigma}\Bigg\langle m(1+x)\left(e^{-ix^2\sigma g_s^2}-e^{-ix^2\sigma g_u^2}\right)\\
&+e^{-ix^2\sigma g_s^2}\left\{\hat{\pi}_{p'}(\theta'_{0,\sigma})-\hat{\pi}_{p'}(\phi)+x\int_0^1d\eta\,[\hat{\pi}_{p'}(\theta'_{0,\eta\sigma})-\hat{\pi}_{p'}(\phi)]\right.\\
&\quad\left.+\frac{x}{1-x}\frac{\hat{n}\hat{\Delta}_{0,s}\hat{\pi}_{p'}(\theta'_{0,\sigma})}{2p'_-}-\frac{x(2-x)}{1-x}\frac{\hat{n}\hat{\Delta}_{0,s}\hat{g}_s}{2p'_-}-2x^3\sigma(g_sg_{1,s})\hat{n}\right\}\\
&-e^{-ix^2\sigma g_u^2}\left\{\hat{\pi}_p(\theta_{0,\sigma})-\hat{\pi}_p(\phi)+x\int_0^1d\eta\,[\hat{\pi}_p(\theta_{0,\eta\sigma})-\hat{\pi}_p(\phi)]\right.\\
&\quad\left.\left.-\frac{x}{1-x}\frac{\hat{\pi}_p(\theta_{0,\sigma})\hat{n}\hat{\Delta}_{0,u}}{2p_-}+\frac{x(2-x)}{1-x}\frac{\hat{g}_u\hat{n}\hat{\Delta}_{0,u}}{2p_-}-2x^3\sigma(g_ug_{1,u})\hat{n}\right\}\right\rangle.
\end{split}
\end{equation}
This expression also shows that $\Gamma_{R,q}(p,p',q;\phi)$ tends to zero for $A^{\mu}(\phi)\to 0$.

\subsection{Some considerations about the LCFA}

As we have mentioned in the main text, in order to study the structure component $\Gamma_{R,q}^{\mu}(p,p',q;\phi)$ of the vertex-correction function, it is useful to rewrite the phases $G_s^2/(s+t)$ and $G_u^2/(u+t)$ in a convenient form [see Eq. (\ref{Gamma_q})]. Indeed, since, as we have seen in the main text, the component $\Gamma_{R,q}(p,p',q;\phi)$ does not contribute to any transition matrix element, we only report here some considerations about these phases. By starting from the identity $v^2=2v_+v_--\bm{v}^2_{\perp}$, valid for a generic four-vector $v^{\mu}$, it can easily be shown that
\begin{align}
\label{G_s^2}
G_s^2&=s^2(m^2+\delta m_{0,s}^2),\\ 
\label{G_u^2}
G_u^2&=u^2(m^2+\delta m_{0,u}^2),
\end{align}
where we have introduced the laser-induced square mass corrections
\begin{align}
\label{m_corr_0s}
\delta m_{0,s}^2&=\frac{1}{s}\int_0^sds'\bm{\mathcal{A}}^2_{\perp}(\psi_{0,s'})-\frac{1}{s^2}\left[\int_0^sds'\bm{\mathcal{A}}_{\perp}(\psi_{0,s'})\right]^2=\frac{1}{s}\int_0^sds'\bm{\Delta}^2_{0,s',\perp}-\frac{1}{s^2}\left(\int_0^sds'\bm{\Delta}_{0,s',\perp}\right)^2,\\
\label{m_corr_0u}
\delta m_{0,u}^2&=\frac{1}{u}\int_0^udu'\bm{\mathcal{A}}^2_{\perp}(\psi_{0,u'})-\frac{1}{u^2}\left[\int_0^udu'\bm{\mathcal{A}}_{\perp}(\psi_{0,u'})\right]^2=\frac{1}{u}\int_0^udu'\bm{\Delta}^2_{0,u',\perp}-\frac{1}{u^2}\left(\int_0^udu'\bm{\Delta}_{0,u',\perp}\right)^2.
\end{align}
We notice that for the present case of on-shell electrons, the quantities $G_s^2$ and $G_u^2$ are non-negative. By proceeding as in the main text we arrive to the approximated expressions
\begin{align}
\label{m_corr_0_LCFA}
\delta m_{0,s}^2&\approx \frac{1}{3}m^2\left(\frac{t}{s+t}\right)^2m^4s^2\chi_{p'}^2(\phi),&& \delta m_{0,u}^2\approx \frac{1}{3}m^2\left(\frac{t}{u+t}\right)^2m^4u^2\chi_p^2(\phi),
\end{align}
where $\chi_p(\phi)=p_-|\bm{\mathcal{E}}_{\perp}(\phi)|/m^3$ and $\chi_{p'}(\phi)=p'_-|\bm{\mathcal{E}_{\perp}}(\phi)|/m^3$, with $\bm{\mathcal{E}}_{\perp}(\phi)=-\bm{\mathcal{A}}'_{\perp}(\phi)$. These approximated expressions are valid if $\eta_0/\chi_0^{2/3}=\chi_0^{1/3}/\xi_0\ll 1$ \cite{Baier_1989,Khokonov_2002,Di_Piazza_2007,Dinu_2016,Di_Piazza_2018_c,Di_Piazza_2019,Ilderton_2019_b,Podszus_2019,Ilderton_2019}. The final expressions of the phases $G_s^2/(s+t)$ and $G_u^2/(u+t)$ within the LCFA are
\begin{align}
\frac{G_s^2}{s+t}&=\frac{s}{s+t}m^2s\left[1+\frac{1}{3}\frac{t^2}{(s+t)^2}m^4s^2\chi_{p'}^2(\phi)\right],\\ 
\frac{G_u^2}{u+t}&=\frac{u}{u+t}m^2u\left[1+\frac{1}{3}\frac{t^2}{(u+t)^2}m^4u^2\chi_p^2(\phi)\right].
\end{align}
The above expressions simplify by means of the mentioned changes of variables:
\begin{align}
\label{Phase_0s_LCFA}
\frac{G_s^2}{s+t}&=x^2m^2\sigma\left[1+\frac{1}{3}x^2(1-x)^2m^4\sigma^2\chi_{p'}^2(\phi)\right],\\ 
\label{Phase_0u_LCFA}
\frac{G_u^2}{u+t}&=x^2m^2\sigma\left[1+\frac{1}{3}x^2(1-x)^2m^4\sigma^2\chi_p^2(\phi)\right].
\end{align}

\end{document}